\UseRawInputEncoding
\PassOptionsToPackage{table}{xcolor}
\documentclass[%
 reprint,
 superscriptaddress,
 amsmath,amssymb,
 aps,
 prresearch,
floatfix,
]{revtex4-2}

\usepackage{tikz}

\usetikzlibrary{positioning,fit}
\usetikzlibrary{quantikz2}
\usepackage{graphicx}
\usepackage{dcolumn}
\usepackage{bookmark}
\usepackage{bm}
\usepackage{braket}
\usepackage{tabularx}
\usepackage{xcolor}
\usepackage{threeparttable}
\usepackage{graphicx}
\usepackage{mathtools}
\usepackage{multirow}

\usepackage[caption=false]{subfig}
\usepackage{physics}
\usepackage{mathrsfs}
\usepackage{xifthen}
\usepackage{array}[=2016-10-06]
\usepackage{enumitem}
\usepackage{graphicx}

\usepackage[utf8]{inputenc}
\usetikzlibrary{positioning} 
\usepackage[caption=false]{subfig}

\usepackage{graphicx,amsmath,amssymb,tikz}
\usetikzlibrary{positioning}

\newcommand{\processor}{\texttt{Nighthawk R2}}

\newcommand{\pnl}{Physical and Computational Sciences, Pacific Northwest National Laboratory, Richland, Washington, 99354, USA}
\newcommand{\uw}{Department of Electrical and Computer Engineering, University of Washington, Seattle, Washington, 98195, USA}

\begin{document}

\preprint{APS/123-QED}

\title{Early Fault-Tolerant Magic State Synthesis and Spacetime Efficient Analog Rotations with Color Code on a Superconducting Quantum Processor}

\author{Sean Garner\textsuperscript{\S}}
\email{seangarn@uw.edu}
\affiliation{\pnl}
\affiliation{\uw}

\author{Aaron Hoyt\textsuperscript{\S}}
\email{ahoyt@uw.edu}
\affiliation{\pnl}
\affiliation{Department of Physics, University of Washington,
Seattle, Washington, 98195, USA}

\author{Shuwen Kan}
\affiliation{\pnl}
\affiliation{Department of Computer Science, Fordham University,
New York City, New York 10458, USA}

\author{Samuel Stein}
\affiliation{\pnl}

\author{Chenxu Liu}
\email{chenxu.liu@pnnl.gov}
\affiliation{\pnl}

\author{Ang Li}
\affiliation{\uw}

\date{\today}

\begin{abstract}
With recent improvements to syndrome extraction scheduling and decoding, and the availability of complete transversal Clifford operations and lattice surgery, color code is an especially promising platform for quantum processors with local connectivity. The key ingredient for universal computation is thus generating useful magic states with limited resources. Space-Time Efficient Analog Rotations (STAR) and Fault-tolerant magic state synthesis (MSS) are two such methods for magic state generation, targeted for near-term and long-term quantum computing, respectively. 
In this work, we perform the first implementations of both magic state paradigms on a superconducting quantum processor. We present a STAR-type injection protocol on the color code, with favorable error rate and spacetime costs versus prior protocols on both the color and surface codes. We then implement a MSS factory by performing multi-patch magic state cultivations, interleaved with transversal Clifford operations, and lattice-surgery magic state teleportation. Under post-selection, we achieve a kickback tomography correct sign probability of $99.9859\%$ on cultivated $\ket{T}_L$ states, $96\%$ tomographic fidelity on logical analog rotation states, and $91\%$ fidelity on synthesized logical magic states consuming 3 $T_L$-states, 3 transversal Hadamard operations,  and real-time feed-forward $S_L$ corrections during magic state teleportation. We compare the efficacy of post-selection, Most-Likely-Error (MLE) decoding, and concatenated-matching decoding throughout, highlighting the potential of end-to-end color code computation.
\end{abstract}

\makeatletter
\appdef\frontmatter@footnote@produce{%
  \begingroup
    \def\@thefnmark{\S}%
    \frontmatter@footnotetext{%
      These authors contributed equally to this work.%
    }%
  \endgroup
}
\makeatother

\maketitle

\section{Introduction}

Quantum computing offers promising speedups for problems that are computationally challenging for classical computers, including integer factorization~\cite{shor_factoring_1997, shor1999polynomial}, quantum simulation of chemical and material systems~\cite{lloyd_simulators_1996, georgescu2014quantum, kandala2017hardware}, unstructured search~\cite{Grover1996}, and linear algebra~\cite{HHL2009}. Realizing these advantages at scales of scientific and technological interest, however, requires reliable quantum computations in the presence of physical noise, including decoherence, relaxation, and imperfections in hardware and control. Quantum error correction (QEC) provides a pathway toward reliable computation by encoding quantum information into logical qubits and repeatedly extracting error syndromes to identify and correct errors~\cite{shor1995, Preskill1998}. 

Recent experimental advances have demonstrated several key building blocks of fault-tolerant quantum computing (FTQC) with different QEC codes across different hardware platforms. Superconducting processors have achieved below-threshold surface-code quantum memories, with logical error rates decreasing as the code distance increases~\cite{google_quantum_ai_and_collaborators_quantum_2024, kan2026, lavasani_quantum_2026}, logical operations implemented through lattice surgery~\cite{wang_lattice_surgery_2026}, and magic state preparation~\cite{rosenfeld_magic_2026}. Neutral-atom systems have demonstrated repeated QEC, magic-state distillation, and universal logical-gate primitives~\cite{Rodriguez2025, Bluvstein2026, Muniz2025, atom_computing2026}, while trapped-ion systems have shown logical Clifford operations that outperform the unencoded physical implementations~\cite{berthusen2026}. These milestones show the emergence of the early FTQC regime, where intermediate-sized QEC can be utilized to protect the quantum information in the computation. However, fully FTQC execution of application-scale quantum algorithms remains far beyond the capabilities of current quantum hardware. For example, the factorization of RSA-2048 still impose stringent demands on hardware scale and overall resource even with deep optimization of the full architecture~\cite{gidney2025factor2048,cain2026shor}. Whether emerging QEC-capable quantum hardware, despite their limited logical-qubit capacity, can show meaningful quantum utility without the overhead of full FTQC becomes an important question.

One of the major sources of overhead in FTQC is the non-Clifford operations. Eastin--Knill theorem asserts that a QEC code cannot support a universal set of transversal logical gates~\cite{Eastin2009}. In widely studied two-dimensional surface-code and color-code architectures, a common approach to implementing non-Clifford gates relies on preparing high-fidelity magic states and consuming them through gate teleportation~\cite{Bravyi2005, nielsen_chuang_2010}. However, preparing these resource states introduces substantial overhead. Magic-state distillation suppresses errors by processing multiple noisy resource states~\cite{Bravyi2005, Litinski2019magicstate}, while magic-state cultivation progressively improves the fidelity of an encoded resource state through error detection and code growth~\cite{gidney2024cultivation}. This overhead is further amplified by the synthesis of arbitrary-angle logical rotations. In the Clifford+$T$ framework, a general logical Pauli-Z rotation $R_{Z}$ must be approximated using a discrete sequence of Clifford and $T$ gates, where ancilla-free synthesis can require tens to hundreds of $T$ gates per rotation, depending on the target approximation precision~\cite{ross_selinger_2016}.

To reduce the cost and locate near-term quantum utilities, Partial Fault-Tolerant (Partial-FT) architectures are proposed to trade complete error protection for reduced resource costs~\cite{akahoshi_partially_2024, yin_flexion_2025,  Dangwal2025, toshio_star-magic_2026, Ismail2026, Yamamoto2026}. An example is the space-time efficient analog rotation (STAR) architecture, which combines FT Clifford operations with low-error, but non-FT, non-Clifford rotations implemented through the injection of encoded resource states~\cite{akahoshi_partially_2024}. By avoiding costly magic-state distillation and the synthesis of arbitrary rotations into Clifford+$T$ gates, STAR offers a promising route toward useful quantum computation with substantially fewer physical resources~\cite{toshio_star-magic_2026}. However, these advantages have largely been assessed through theoretical analysis and numerical simulations. Their realization on quantum hardware remains an important experimental challenge, while establishing the practical performance of STAR-type protocols is essential to determining their viability for near-term error-corrected quantum computation.

With the recent development of Concatenated-MWPM~\cite{lee2025color} and other decoders, 2D color codes are increasingly promising for experimental QEC. Unlike surface codes, color codes support the full single-qubit Clifford group transversally, enabling efficient logical operations that are central to both partial and full FT architectures~\cite{bombin_topological_2006, landahl_fault-tolerant_2011}. Moreover, their structure supports resource-efficient magic-state cultivation protocols that exploit transversal Clifford operations to suppress preparation errors~\cite{gidney2024cultivation}. These properties make color codes attractive for both low-overhead STAR-type injection and high-fidelity magic-state factories. Although recent experiments have demonstrated color-code magic-state injection and cultivation~\cite{lacroix_scaling_2025, rosenfeld2026cultivation}, constructing a hardware-efficient STAR-type injection protocol on color code, and comparing its experimental fidelity and spacetime overhead with conventional magic-state preparation approaches, remain open questions. More broadly, a systematic experimental comparison of these approaches is needed to understand their respective advantages and limitations on near-term quantum hardware.

In this paper, we develop and experimentally benchmark a color-code injection protocol for partially fault-tolerant analog rotations. The protocol combines local two-qubit $ZZ$ rotations with the superdense color-code layout and syndrome-extraction circuit~\cite{gidney_circuits_2023}. Under the uniform circuit-level depolarizing noise model, our fault analysis identifies a single undetected harmful first-order contribution, i.e., a correlated $Z\otimes Z$ fault at the injection gate. After postselection and growth to the target code distance, the conditional logical error probability of an accepted preparation is $P_L=p/15+O(p^2)$, where $p$ is the physical error probability. We implement this protocol alongside an established color-code injection scheme~\cite{zhang_facilitating_2024} on IBM \processor{}, a superconducting quantum processor with square-grid connectivity and mid-circuit measurement capability. Our experiments characterize the performance of direct logical-state preparation and establish a hardware baseline for analog logical rotations. We further integrate magic-state cultivation with logical gate teleportation on the same processor, demonstrating synthesized non-Clifford resource states within the color-code setting. By studying both routes within a common hardware setting, this work connects improvements in physical control and error detection to the accuracy and overhead of logical non-Clifford operations, informing how Partial-FT can make effective use of early QEC-capable processors.

\section{Partial Fault-Tolerant Color Code Injection}

\label{sec:injection}

Partial-fault-tolerance (Partial-FT) is a bridge between bare NISQ operations and end-to-end FTQC. In partial-FT protocols, the most difficult resource states are allowed to have lapses in fault tolerance in exchange for major benefits in resource efficiency. Less resource-intensive logical operations, such as transversal Clifford gates in the color code and surface code, remain fault-tolerant. This combination seeks to make the best tradeoff between both paradigms to accelerate useful applications. The STAR architecture is one such partial-FT design which aims to enable useful quantum algorithms with the surface code~\cite{akahoshi_partially_2024}. We also exploit $R_{ZZ}(\theta)$  operations, natively injection available on \processor~along with the congruency of Steane code stabilizers, to embed a high-quality arbitrary injection directly in the color code. Our protocol allows us to realize a lower error rate than previous color code and surface code injection protocols using one- and two-qubit injection gates~\cite{akahoshi_partially_2024,zhang_facilitating_2024,lacroix_scaling_2025}, and keeps the $R_{ZZ}$ local. This avoids the need for SWAP gates on a square grid processor, which would otherwise sacrifice first-order errors. Like the STAR architecture's [[4,1,1,2]] subsystem code, an intermediate check for scheduling many injection-retries in parallel can be performed simply as a single-color stabilizer check,
\begin{equation}
    S = \{X_0X_1X_2X_3, Z_0Z_1Z_2Z_3\}.
\end{equation}

\subsection{STAR-Type Protocol on the Superdense Color Code}

We assume the superdense color code schedule throughout according to the design of Ref.~\cite{gidney_new_2023}, with two ancilla qubits per color code face. Each face's ancilla qubits are initialized in a Bell pair before syndrome extraction, enabling only weight-4 connectivity and X/Z checks in a single parallelized round of syndrome extraction. These ancilla also serve as intermediaries for preparation in our scheme where needed. We begin by presenting the end-to-end arbitrary rotation injection scheme with the best found analytical logical error rate. 

\subsubsection{Magic state preparation in subsystem code}
The asymptotic advantage of STAR-type injection is the $R_{ZZ}(\theta)$ operation. Multi-qubit rotation has been used in prior works to implement analog multi-qubit rotations~\cite{akahoshi_partially_2024, toshio_practical_2025,toshio_star-magic_2026,Yamamoto2026}. Here we find a schedule to minimize first- and second-order errors compared to previous protocols by convenient superdense color code embeddings and growth flagging. With a 1-qubit injection site, only one type of flag/stabilizer can protect the injection gate while still allowing an arbitrary rotation. With a 2-qubit injection gate, we design stabilizers to detect all 2-qubit Pauli errors aside from $Z\otimes Z$ errors, which must be undetected if analog rotations are to be allowed.

\begin{figure}[!h]
    \centering
    \includegraphics[width=.45\linewidth]{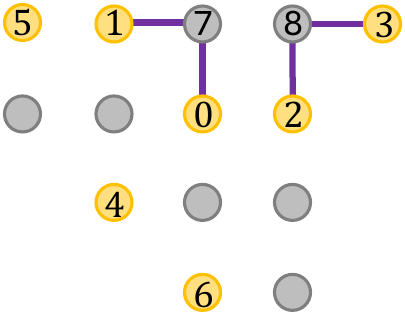}
    \includegraphics[width=.45\linewidth]{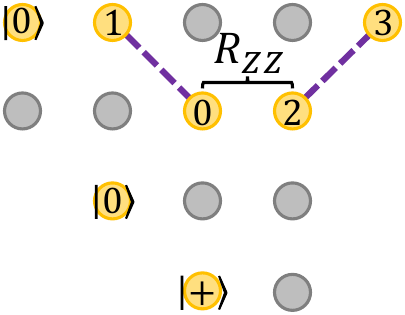}
    \caption{Best found schedule to prepare $R_{ZZ}(\theta)$ injection and state initialization prior to a [[4,1,1,2]] and/or honeycomb color code check. \textbf{Left:} Qubits 0-1 and 2-3 are initialized in their respective Bell states via their nearest ancilla. \textbf{Right:} $R_{ZZ}(\theta)$ is applied locally to qubits 0-2. Qubits 4, 5, and 6 are initialized in $\ket{0}$, $\ket{0}$, and $\ket{+}$ respectively. Qubit 6 has an ancillary qubit to check the $X$-flag in parallel with the [[7,1,3]] initialization. Afterwards, the ancilla is released.}
    \label{fig:bell_prep}
\end{figure}

The first step for the best first-order detection version of our protocol is a Bell state preparation of what will be the logical $X$'s of our data qubit and logical $Z$'s of our gauge qubit, from the view of an intermediate [[4,1,1,2]] code (equivalent to the \textit{red} boundary face of the later [[7,1,3]] check). We specifically assign the data and gauge in this way for local $R_{ZZ}$ access on the $Z$ logical of the data qubit. The data qubit's logical operators are
\begin{equation}
    L_X = X_0X_1 \quad L_Z = Z_0Z_2
\end{equation}
and the gauge operators are
\begin{equation}
    G_X = X_0X_2 \quad G_Z = Z_0Z_1,
\end{equation}
where the qubit indices are labeled in Fig.~\ref{fig:bell_prep}.

\begin{figure}[htbp]
\centering
\begin{quantikz}[row sep=0.45cm, column sep=0.5cm]
\lstick{$d_0:\ \lvert 0\rangle$} & \gate{H} & \ctrl{1} & \qw      & \ctrl{1} & \ctrl{3}
    \gategroup[wires=4,steps=1,style={dashed,rounded corners,inner sep=1pt},
               label style={label position=above,anchor=south,yshift=0.1cm}]{$R_{ZZ}(\theta)$} & \qw \\
\lstick{$a_7:\ \lvert 0\rangle$} & \qw      & \targ{}  & \ctrl{1} & \targ{}  & \qw      & \qw \\
\lstick{$d_1:\ \lvert 0\rangle$} & \qw      & \qw      & \targ{}  & \qw      & \qw      & \qw \\
\lstick{$d_2:\ \lvert 0\rangle$} & \gate{H} & \ctrl{1} & \qw      & \ctrl{1} & \control{} & \qw \\
\lstick{$a_8:\ \lvert 0\rangle$} & \qw      & \targ{}  & \ctrl{1} & \targ{}  & \qw      & \qw \\
\lstick{$d_3:\ \lvert 0\rangle$} & \qw      & \qw      & \targ{}  & \qw      & \qw      & \qw
\end{quantikz}
\caption{Preparation of the 4-qubit stabilizer code with weight-4 connectivity and local $R_{ZZ}$ in the grid. Figure~\ref{fig:bell_prep} shows the 2D overhead layout.}
\label{circ:rzz_prep}
\end{figure}
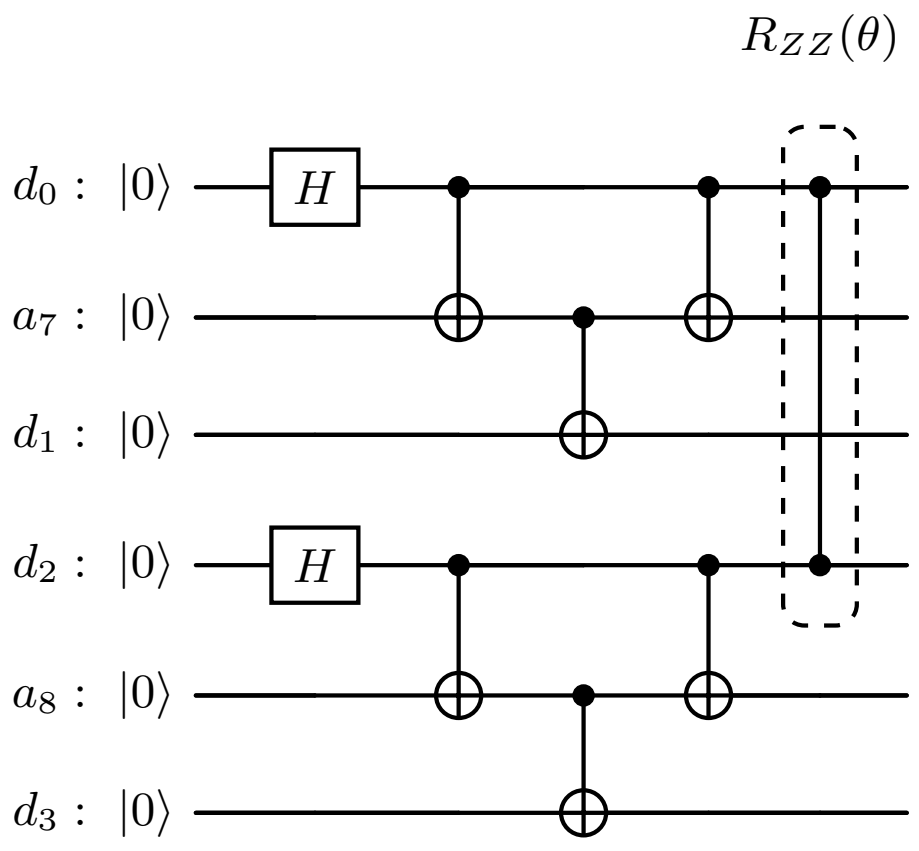

Initializing 0-1 and 2-3 in Bell pairs is not necessary for the end logical state, but it allows us to deterministically check $Z_0Z_1$ and $Z_2Z_3$, protecting against $X$ and $Y$-type combinations of 2-qubit depolarization at the injection gate. Furthermore, the first $XXXX$ check will anti-commute with $IZ$ and $ZI$ type errors at the $R_{ZZ}$ site. Different from the scheme used in Ref.~\cite{Yamamoto2026}, the key insight in this protocol is local $R_{ZZ}$ within the weight-4 $X$ and $Z$ stabilizer space, which both generates a magic state $\frac{1}{\sqrt{2}}(\ket{0} + e^{-i \theta} \ket{1})$ up to global phase, and can encodes into the [[7,1,3]] codespace without routing or opportunity for other first order errors beyond $Z\otimes Z$. The corresponding quantum circuit is shown in Fig.~\ref{circ:rzz_prep}.

\subsubsection{Color Code Projection}

\begin{figure}[h]
    \centering
    \includegraphics[width=.45\linewidth]{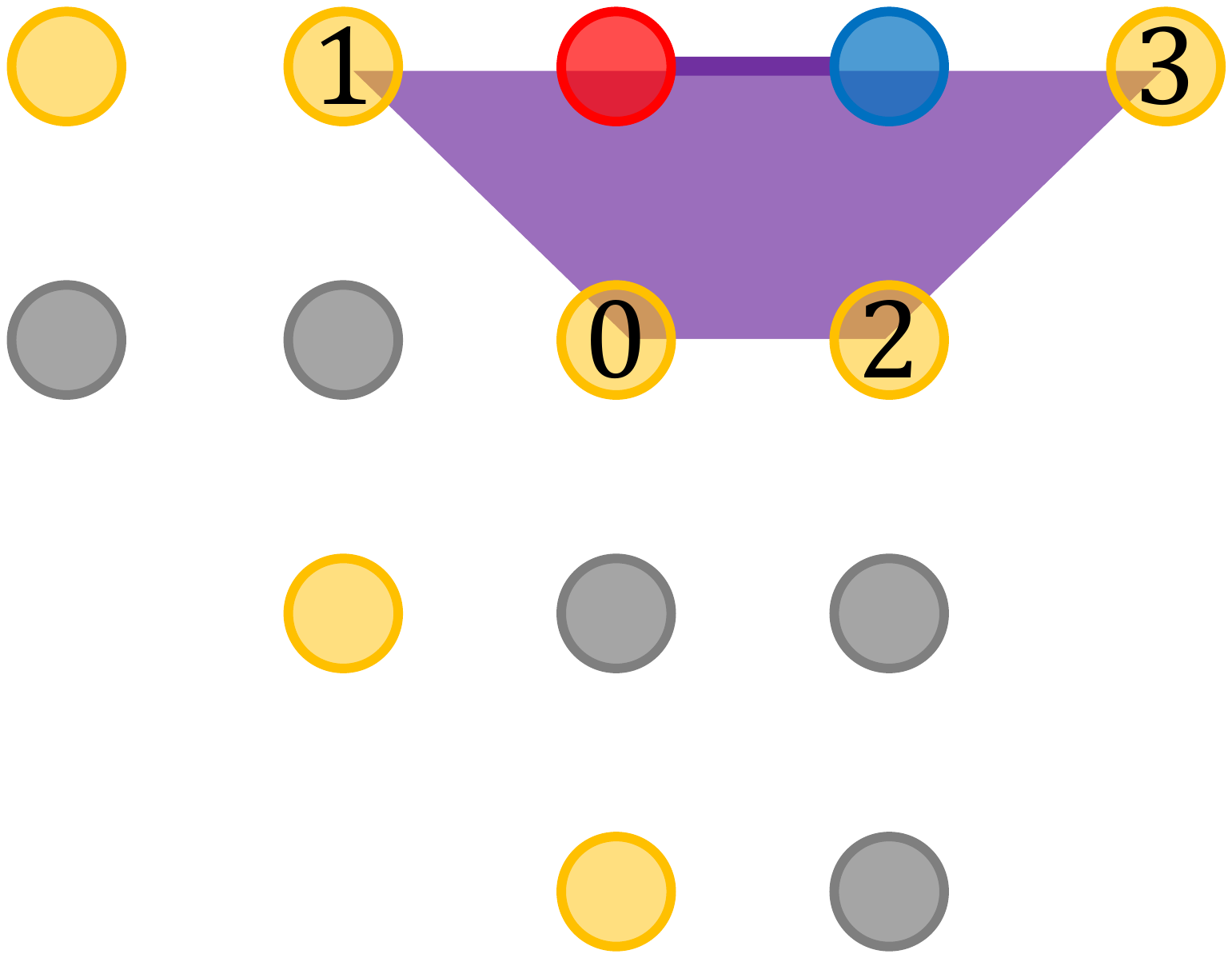}
    \includegraphics[width=.45\linewidth]{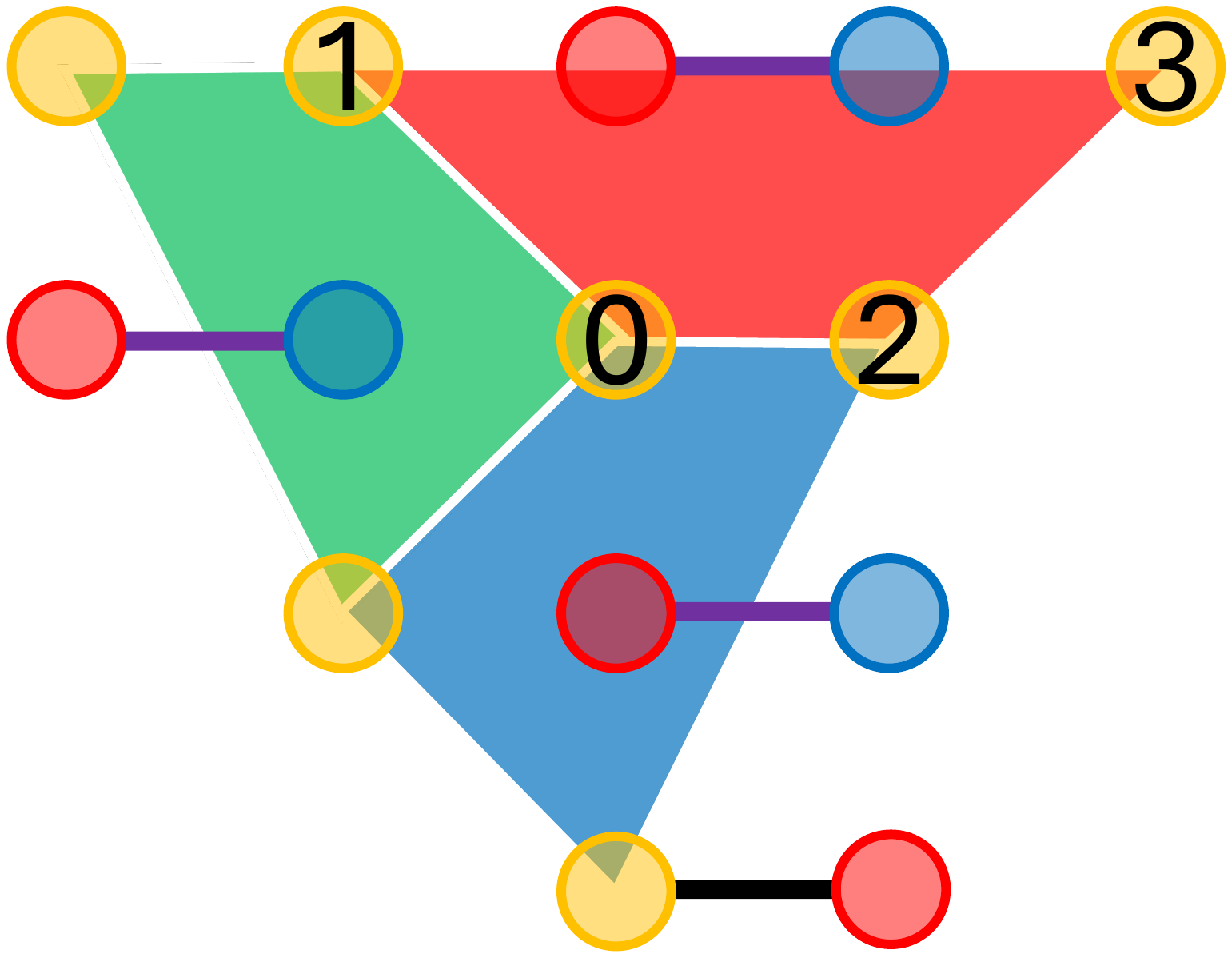}
    \caption{Red and blue ancilla represent $X$ and $Z$ syndrome measurements respectively. Purple lines indicate a Bell pair initialization. \textbf{Left:} Optional [[4,1,1,2]] subsystem check for error detection and retry with fewer qubits. The purple face represents shared $XXXX$ and $ZZZZ$ stabilizers. While this leaves two degrees of freedom for qubits, the logical qubit is the one supported by $Z_0Z_2$, while the other is chosen to be a gauge qubit (where errors are harmless). \textbf{Right:} Colored faces indicate the typical coloring of a 2D color code. For the first round: Red checks are deterministic for both $X$ and $Z$. Green checks are only deterministic only for $Z$. Qubit 6 is $X$-flagged in parallel with the non-deterministic blue checks, catching 1q $Z$ errors.}
    \label{fig:steane_code}
\end{figure}

Immediately after the steps in Fig.~\ref{fig:bell_prep}, or after the [[4,1,1,2]] check if desired, the logical state can be projected into the full 2D color code by measuring all stabilizers of the color code. Preparation in Fig.~\ref{fig:bell_prep} was done to maximize the value of deterministic stabilizers in the grown code while still extending the logical operators. In the very first round of projection, using the right side of Fig.~\ref{fig:steane_code} as reference, \textit{red} $X$, \textit{red} $Z$, and \textit{green} $Z$ are the deterministic checks. $YY$ on 0-2 is thus detected by the \textit{green}-$Z$ check. Combined with the $X_0X_1X_2X_3$ and $Z_0Z_1Z_2Z_3$ check of the \textit{red} face, this provides an injection logical error rate of only $P_L={p/15}+O(p^2)$, i.e., $ZZ$ faults on the $R_{ZZ}$ gate, which turns out to be the only first-order logical error in the end-to-end protocol assuming full-context decoding or good logical readout.

Since qubits 5 and 6 must be $\ket{0}$ and $\ket{+}$, respectively, and qubits 0-2 must allow the arbitrary $Z$ rotation, the green $X$ and both $X$ and $Z$ blue checks cannot be deterministic on the first projection. Luckily, green $X$ will not have uncaught $Z$ faults, since 0-1 are captured by the stabilizer checks marked as red, and qubits 4 and 5 are $\ket{0}$. Qubit 4's value does not matter for the logical operators, but $\ket{0}$ turns out to be a good initialization to allow the previously discussed $Y_0\otimes Y_2$ catch.

Qubit 6 then is an obvious spot for first-order errors, as the blue face is entirely non-deterministic and we required a Hadamard operation to extend the logical operator (assuming \processor's basis gates). This makes qubit 6 a great candidate for an $X$ flag check, using one ancilla in the square grid which is otherwise unused in the superdense [[7,1,3]] code. Performing this flag in parallel with the Steane code projection is thus sufficient to remove the first-order errors that would come from a one-qubit Pauli error around the initialization Hadamard operation.

\begin{figure}[h]
    \centering
    \includegraphics[width= 0.9 \linewidth]{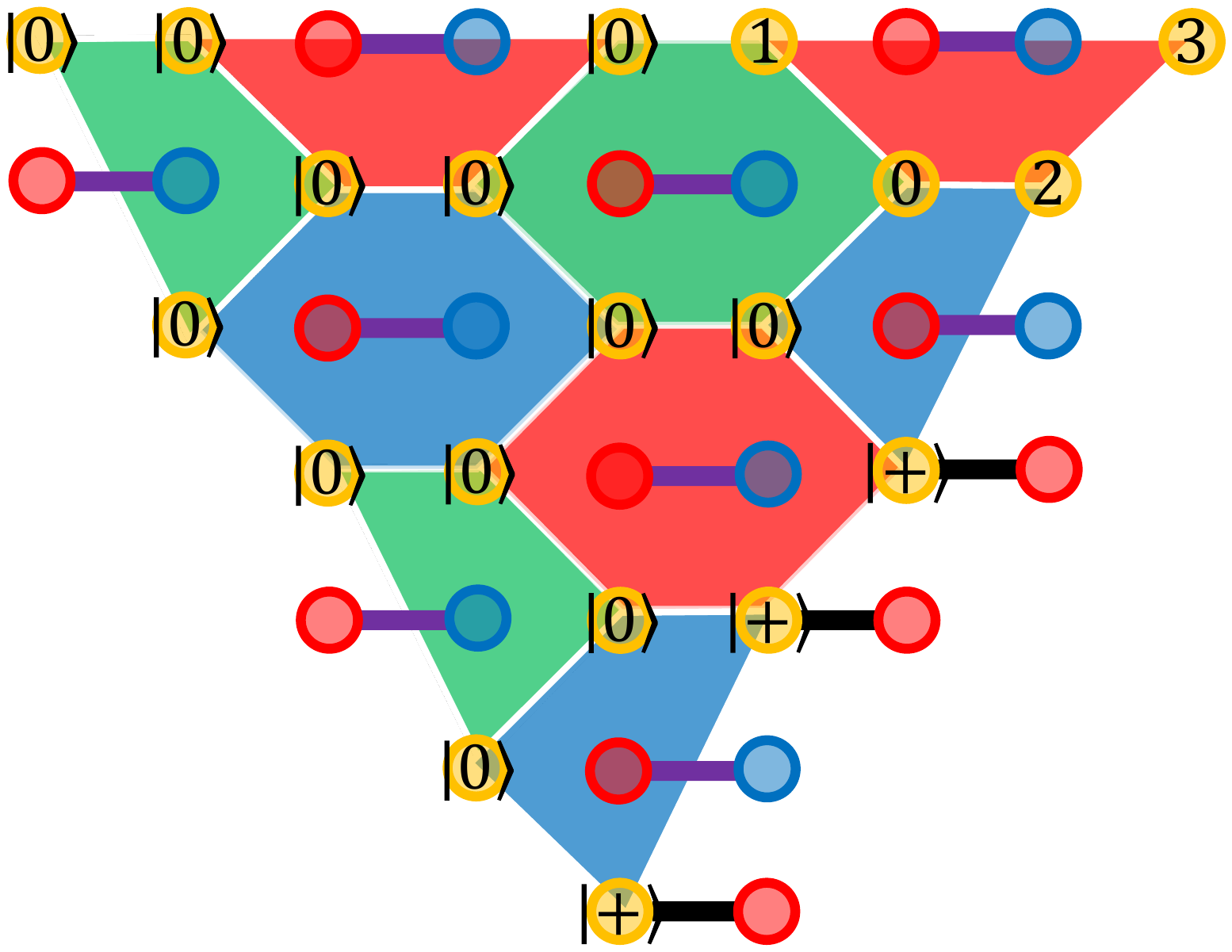}
    \caption{Square grid layout of the distance-5 honeycomb color code with superdense stabilizer ancilla and $X$ flags on the logical-$X$ extension qubits.}
    \label{fig:d5_code}
\end{figure}

The same first-order logical error for the distance-3 case can be recovered for higher-distance color codes by extending the logical $X$ and $Z$ observables, extending the inner $\ket{0}$ initializations for deterministic $Z$ checks, and adding flag qubits to the logical $X$ boundary. Fig.~\ref{fig:d5_code} shows the full initialization and syndrome check pattern.

\subsection{Protocol comparison with circuit noise}

The straightforward benefits of our protocol versus prior protocols for arbitrary state injection on planar codes, is the usage of a native multi-qubit rotation gate for injection, the congruency of [[4,1,1,2]] color code stabilizers, and the timing of the flag qubit. Multi-qubit rotation makes the undetectable logical error configuration geometrically unlikely on the injection gate. The existing weight-4 checks of color code boundaries also naturally support a logical projection of the encoding rotation without the explicit need for an intermediate check, and careful flagging removes the remaining gate-based first-order error, while using less resources than a surface code.

\begin{figure*}[t]
    \centering
    \subfloat[]{%
        \includegraphics[width=0.45\textwidth]
            {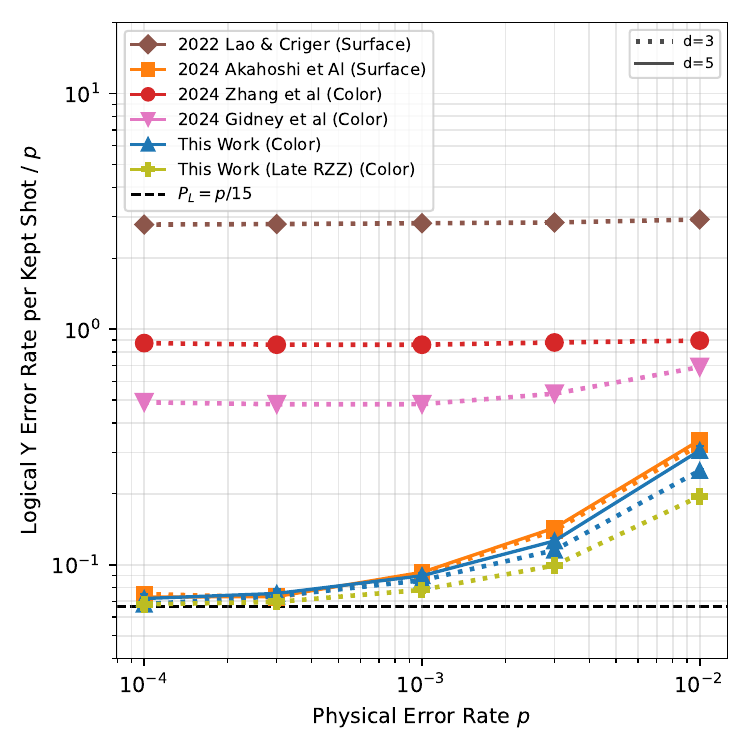}%
        \label{fig:protocol_simulation}%
    }
    \hfill
    \subfloat[]{%
        \includegraphics[width=0.45\textwidth]
            {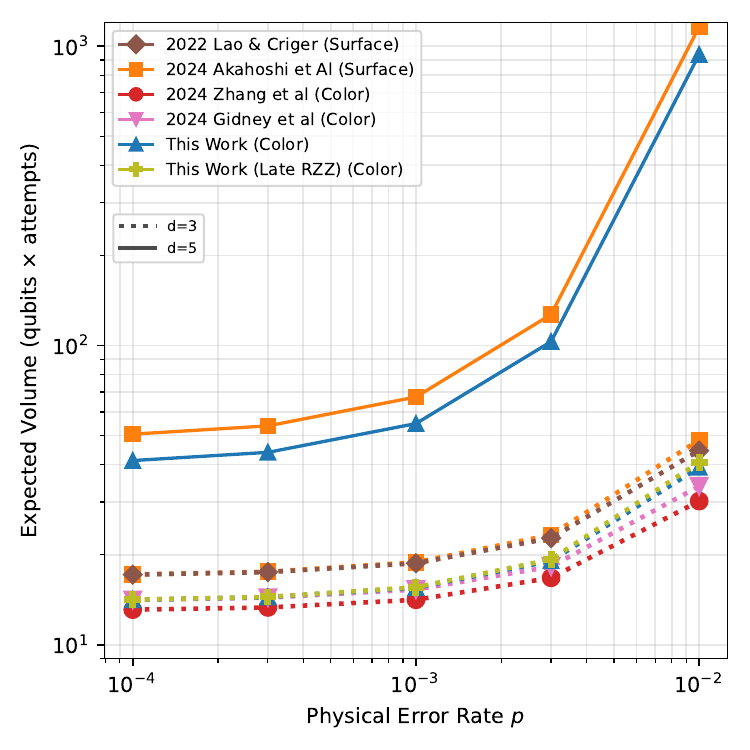}%
        \label{fig:protocol_spacetime}%
    }
    \caption{
        Comparison of logical error rates and resource overheads for magic-state injection using surface and color codes.
        \textbf{(a)} Simulated logical $Y_L$ error for $\ket{S}$ injection using different protocols. Sweeping $\theta$ at other rotation angles does not affect the simulated infidelity, so the $S$ gate is chosen for deterministic logical-error analysis. \texttt{This Work (Color)} corresponds to the protocol in Figs.~\ref{fig:bell_prep} and~\ref{fig:steane_code}. \texttt{Late RZZ} applies $R_{ZZ}(\theta)$ after a $[[4,1,1,2]]$ check, initializing the code for a deterministic logical $R_{ZZ}$ operation. The intermediate check suppresses certain higher-order logical errors arising from combinations of Bell-state preparation and $R_{ZZ}$ gate errors. The tradeoff is delayed post-selection for certain errors, such as isolated $R_{ZZ}$ errors that are not detected until the $[[7,1,3]]$ check.
        \textbf{(b)} Product of the number of occupied qubits and the
        expected number of attempts required to obtain an accepted state through the complete injection protocol, up to the indicated code distance. Despite the additional ancillas required for superdense syndrome extraction, the color-code protocol achieves lower spacetime volume throughout, assuming parallelized measurements within each QEC round and no intermediate checks for any method. Both surface-code STAR and our protocol support an intermediate $[[4,1,1,2]]$ code, but our protocol requires fewer qubits and two-qubit gates for code preparation and the local $R_{ZZ}$ operation.
    }
    \label{fig:protocol_comparison}
\end{figure*}

Fig.~\ref{fig:protocol_simulation} shows how prior injection and deterministic check protocols (Refs.~\cite{lao_magic_2022},~\cite{zhang_facilitating_2024}) lag partly-stabilized protocols (Refs.~\cite{gidney_magic_2024},~\cite{akahoshi_partially_2024}) because they allow truly arbitrary states, rather than arbitrary rotation in a plane. Surface code STAR and our STAR-type color protocol then achieve a reduced order of magnitude in a uniform circuit-level noise model largely by taking advantage of the $p/15$ versus $2p/3$ first order injection error permutations. Our work achieves the remaining benefit over STAR by providing a local embedding for the $R_{ZZ}$ rather than via SWAP gates, and by finding a convenient preparation-$R_{ZZ}$-projection with fewer opportunities for initialization error before the code is protected at distance-3.

The color code protocol, even with superdense ancilla and flagging, is able to use fewer qubits for the same code distance, although this advantage is only constant rather than a better encoding rate. This difference accounts for most of the spacetime volume improvement in Fig.~\ref{fig:protocol_spacetime}. Extra post-selections at high-error rate account for the slightly higher volumes of the 2-qubit gate injections versus 1-qubit gate protocols, although this only shows up at higher error. The 2-qubit gate protocols are also shown with deterministic distance-5 injection to show how this volume scales with distance and error rate.

\section{Fault-tolerant Color Code Magic State Factory}

\subsection{Magic State Synthesis}
2D color code supports a transversal implementation of the entire logical Clifford group~\cite{bombin_topological_2006,landahl_fault-tolerant_2011}. By Eastin--Knill theorem, it cannot support any transversal non-Clifford operation~\cite{Eastin2009}. Thus, implementing magic states with quality that matches the transversal Clifford operations requires significant resources, often by starting with a state injection, followed by checks, double-checks, and logical state distillation.

Since magic states on planar codes are a critical bottleneck, a frequently proposed component of fault-tolerant quantum computers is magic state factories. Magic state factories are generally patches of qubits with the primary purpose of providing high-quality magic states with sufficient throughput to the computation. The checks and double checks required to obtain the desired state fidelity require frequent throwaways and retries, so separating this out from the entire computation prevents the waste of other resources, and ensures magic states can be queued and delivered in time such that the rest of the computation does not idle. Separating out the factory also provides space for distillation, where many decent magic states must be distilled to make a suitable resource state for fault-tolerant computations.

In this work, we specifically investigate and implement three critical steps of the magic state factory: state injection, magic state cultivation, and magic state teleportation. These steps push the limit of what is possible with the number of qubits available today. Implementing the factory needed for FTQC will likely need a growth or escape from the cultivated code to a higher distance code, in addition to a distillation of these high distance codes, once a sufficient amount of high-distance codes are prepared and idling. These two stages require scaling to thousands of qubits to even achieve a single factory.

\begin{figure}[h]
    \centering
    \includegraphics[width=\linewidth]{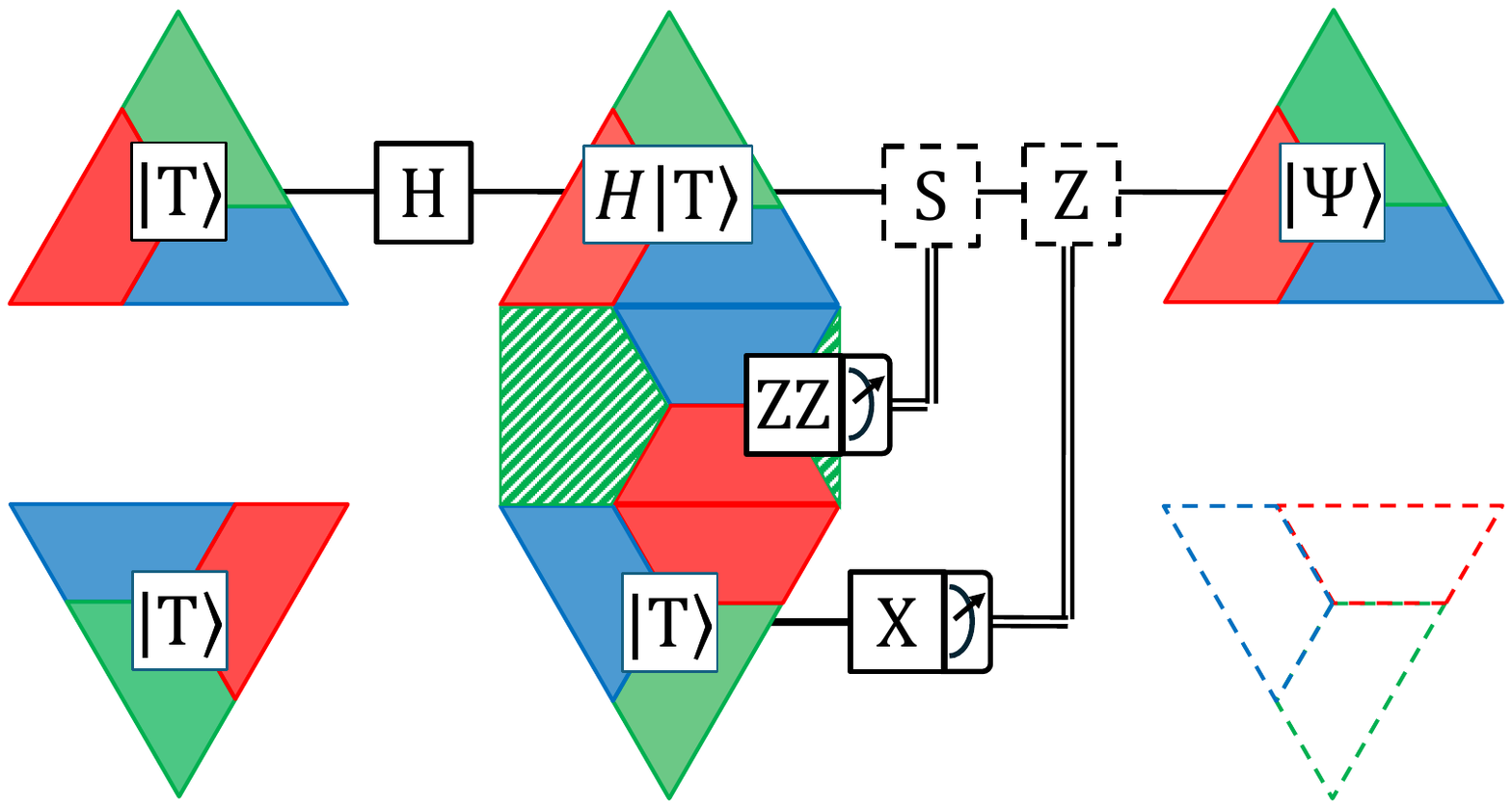}
    \caption{High-level overview of the two-logical-qubit rotation synthesis pipeline. $\ket{T}$ states are injected and cultivated in parallel, and a transversal Hadamard is applied to the 'compute' patch. Both patches are connected via lattice surgery, and the logical $ZZ$ parity of both is measured. Feed-forward $S$ is applied depending on the value of $ZZ$, which tells us whether $T$ or $T^\dagger$ was teleported. The 'factory' patch is then measured out in the $X$ basis to determine a $Z$ correction. In a scalable fault-tolerant architecture, $S$ must be applied in order since it does not commute trivially through the preceding computation in the way that a Pauli correction does. We implement both online and offline corrections in Section~\ref{sec:demo}.}
    \label{fig:synthesis_factory}
\end{figure}

\subsubsection{Magic State Injection}
Implementing any magic state in a planar code is the first challenge. In simulation and implementation, we test a number of different methods for injecting a $\ket{T}$ magic state. We find that, surprisingly at first glance, the best state injection protocol for Partial-FT is not necessarily the best protocol to seed a magic state factory. 

The reason for the gap in protocols turns out to be straightforward. In all of these circuits, the bottleneck is almost entirely the cultivation stage. The cultivation stage will catch and discard almost any malignant error combination at the measurement of the $H_{XY,L}=\frac{1}{\sqrt{2}}(X_L+Y_L)$ observable, since the eigenstates of $H_{XY,L}$ are exactly $\ket{T}_L$ and $Z\ket{T}_L$. Higher injection acceptance can improve the final conditional logical error rate (LER) if the accepted attempts are predominantly harmless after cultivation, as the LER is defined as 
\begin{equation}
    r_{L} = \frac{N_{\text{LE, acc}}}{N_{\text{acc}}},
\end{equation}
where $r_{L}$ is the LER, $N_{\text{acc}}$ is the acceptance count, while $N_{\text{LE, acc}}$ is the number of logical errors in the accepted attempts.

In other words, strong injection protocols remove lots of malignant and benign injection errors, but the malignant errors would otherwise have been caught by cultivation. So, stronger protocols for partial-FT can have the net effect of throwing away otherwise plausible attempts. In our later magic state cultivation, we find that unitary injection-type circuits work best for seeding the cultivated magic state.

\subsubsection{Magic State Cultivation}

The first critical gain that allows $T$-states in particular to achieve higher fidelity than arbitrary $R_Z(\theta)$ states, is magic state cultivation (MSC)~\cite{gidney_magic_2024,rosenfeld_magic_2026}. MSC takes advantage of the fact that the circuit needed to prepare a logical measurement of the $H_{XY,L}$ operator, and thus differentiate between its eigenstates $\ket{T}_L$ and $Z\ket{T}_L$, can be prepared very cheaply on color codes via Clifford operations and a basis change around an $X$ measurement, i.e. $TXT^\dagger$. The only opportunities for logical errors after cultivation are high-order, for example, a prior logical error in the original codespace $\mathcal{C}$ followed by a measurement error during the check, or multiple errors in the short span between the $H_{XY,L}$ measurement and the return to the protected $\mathcal{C}$.

\begin{figure}
\begin{quantikz}[row sep=0.2cm, column sep=0.2cm]
\lstick{$q_0$}            & \gate{T^\dagger} & \targ{}   & \qw       & \qw       & \qw       & \qw       & \qw            & \qw       & \qw       & \qw       & \targ{}   & \gate{T} \\
\lstick{$\ket{+}$}    & \qw              & \ctrl{-1} & \targ{}   & \qw       & \qw       & \qw       & \qw            & \qw       & \qw       & \targ{}   & \ctrl{-1} & \qw \\
\lstick{$q_1$}            & \gate{T^\dagger} & \targ{}   & \ctrl{-1} & \targ{}   & \qw       & \qw       & \qw            & \qw       & \targ{}   & \ctrl{-1} & \targ{}   & \gate{T} \\
\lstick{$\ket{+}$}    & \qw              & \ctrl{-1} & \qw       & \qw       & \qw       & \qw       & \qw            & \qw       & \qw       & \qw       & \ctrl{-1} & \qw \\
\lstick{$\ket{+}$}    & \qw              & \ctrl{1}  & \qw       & \qw       & \qw       & \qw       & \qw            & \qw       & \qw       & \qw       & \ctrl{1}  & \qw \\
\lstick{$q_2$}            & \gate{T^\dagger} & \targ{}   & \ctrl{1}  & \ctrl{-3} & \ctrl{3}  & \meter{X,\ket{+}} & \ctrl{3}  & \ctrl{-3} & \ctrl{1}  & \targ{}   & \gate{T} \\
\lstick{$\ket{+}$}    & \qw              & \ctrl{1}  & \targ{}   & \qw       & \qw       & \qw       & \qw            & \qw       & \qw       & \targ{}   & \ctrl{1}  & \qw \\
\lstick{$q_3$}            & \gate{T^\dagger} & \targ{}   & \qw       & \qw       & \qw       & \qw       & \qw            & \qw       & \qw       & \qw       & \targ{}   & \gate{T} \\
\lstick{$\ket{+}$}    & \qw              & \ctrl{1}  & \ctrl{3}  & \ctrl{2}  & \targ{}   & \qw       & \qw            & \targ{}   & \ctrl{2}  & \ctrl{3}  & \ctrl{1}  & \qw \\
\lstick{$q_4$}            & \gate{T^\dagger} & \targ{}   & \qw       & \qw       & \qw       & \qw       & \qw            & \qw       & \qw       & \qw       & \targ{}   & \gate{T} \\
\lstick{$q_{5}$}         & \gate{T^\dagger} & \qw       & \qw       & \targ{}   & \qw       & \qw       & \qw            & \qw       & \targ{}   & \qw       & \qw       & \gate{T} \\
\lstick{$\ket{+}$} & \qw              & \ctrl{1}  & \targ{}   & \qw       & \qw       & \qw       & \qw            & \qw       & \qw       & \targ{}   & \ctrl{1}  & \qw \\
\lstick{$q_{6}$}         & \gate{T^\dagger} & \targ{}   & \qw       & \qw       & \qw       & \qw       & \qw            & \qw       & \qw       & \qw       & \targ{}   & \gate{T}
\end{quantikz}
\label{fig:cultivation_circuit}
\caption{Measurement of the $H_{XY}$ operator via conjugated $X$ measurement, $H_{XY} = TXT^\dagger$, on a [[7,1,3]] code. Numbered qubits are the data qubits which comprise the Steane code, while all ancilla are prepared in $\ket{+}$ to ensure the $H_{XY}$ check is deterministic. The measured qubit is re-prepared in $\ket{+}$ after measurement}
\end{figure}
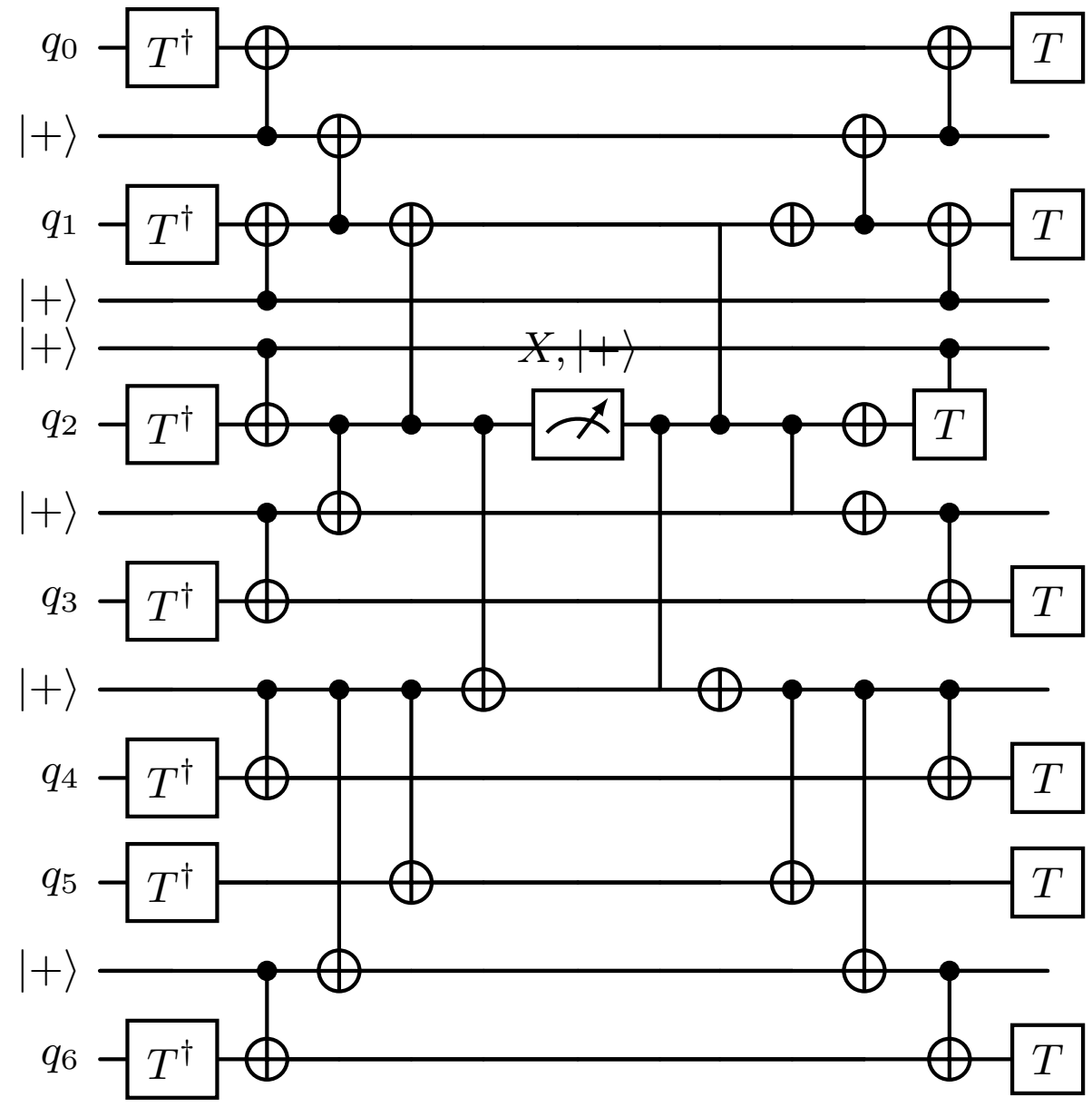

Cultivation can be done for any distance of color code, however, the throwaways increase greatly beyond small distances using post-selection. Error correction is not especially feasible here, since a single measurement error at the $H_{XY,L}$ measurement could give a false negative, which would be a logical error when ``corrected'', binding the error rate to that first-order error.

\subsubsection{Magic State Teleportation}

Magic state teleportation is required to transfer the magic state from the factory to the ongoing computation. Throughout this work, we perform lattice surgery between two color code patches and measure the $Z_{L,C}\otimes Z_{L,F}$ (\textit{compute} and \textit{factory}) parity to teleport the magic state. Taking $\ket{\Psi}_L$ as the ongoing computation state, teleporting $\ket{T}_L$, and choosing $ZZ$ as the parity basis,
\begin{equation}
\ket{\Psi}_{C}\ket{T}_{F}
\xrightarrow{\;M_{ZZ},\,M_{X_F}\;}
\begin{cases}
Z_L^{m_X}T_L\ket{\Psi}_{C},
& (p=1/2),\\[2pt]
Z_L^{m_X}T_L^\dagger\ket{\Psi}_{C},
& (p=1/2).
\end{cases}
\end{equation}
applies $T_L$ up to conditional corrections $S_L$ and $Z_L$. In a noiseless setting, the outcome of $M_{ZZ}$ will be either the +1 or -1 eigenstate branch with equal probability. After the corrections,
\begin{equation}
    Z_L^{m_X}S_L^{-m_{ZZ}}T_L|\Psi\rangle_C\xrightarrow{\;S_L^{{ZZ}}Z_L^{_X}\;}T_L\ket{\Psi}_C.
\end{equation}
In our implementations, we perform magic state synthesis with layered logical operators, $\ket{\Psi}_L = H_L^nT_L^m\ket{+}_L$, and process tomography of magic state teleportation, where $\ket{\Psi}_L \in \{\ket{0}_L,\ket{1}_L,\ket{+}_L,\ket{i}_L\}$.

Magic state teleportation is itself a critical bottleneck to fault-tolerant computation largely because of the decoding requirements. In this work, we have the ability to apply the $S_L$ correction after the computation because of the small computation size considered. However, in a fault-tolerant scheme, this $S$ correction is generally highly sensitive to decoding latency, since whether or not to apply the correction can depend on errors during lattice surgery. The decoder will determine if the syndromes in the surgery region, which have a deterministic parity in the noiseless case, were correct. If not, the decoder must make the correction such that the correct choice of  $S$ or $I$ is made, or else the teleportation is a source of uncorrected logical error. Since $S$ is Clifford, this correction does not trivially propagate through the rest of the computation like Pauli frame updates, so it should be done just-in-time to prevent backlog. However, if the decoder is slower than the next operation, often the case on solid state processors, the computation will have to perform idling syndrome extraction to catch decoherence, which itself creates more decoder work~\cite{terhal_quantum_2015}. 

\subsection{Rotation synthesis}

Combining injection, cultivation, and teleportation with transversal Clifford gates is sufficient to quickly synthesize arbitrary logical rotations (Solovay-Kitaev theorem~\cite{kitaev_quantum_1997,dawson_solovay-kitaev_2005}) on color code up to some error $\epsilon$ using the no-ancilla Ross-Selinger synthesis algorithm (Gridsynth)~\cite{ross_optimal_2016}. Larger fault-tolerant systems will likely grow and distill these states further, but we can push current synthesis by cultivating multiple T-states in parallel on small codes.

\section{Implementation}

\label{sec:demo}

We implement partial-FT injection, magic state cultivation, and gate teleportation on \texttt{ibm\_phoenix}, a~\processor~superconducting processor with square grid connectivity. Logical qubits occupy color code patches with $d=3$, using superdense syndrome extraction with two ancillas prepared as one Bell pair per stabilizer face~\cite{gidney_new_2023}.

\begin{figure}[!tbp]
\centering
\includegraphics[width=\linewidth,height=0.69\textheight,keepaspectratio]{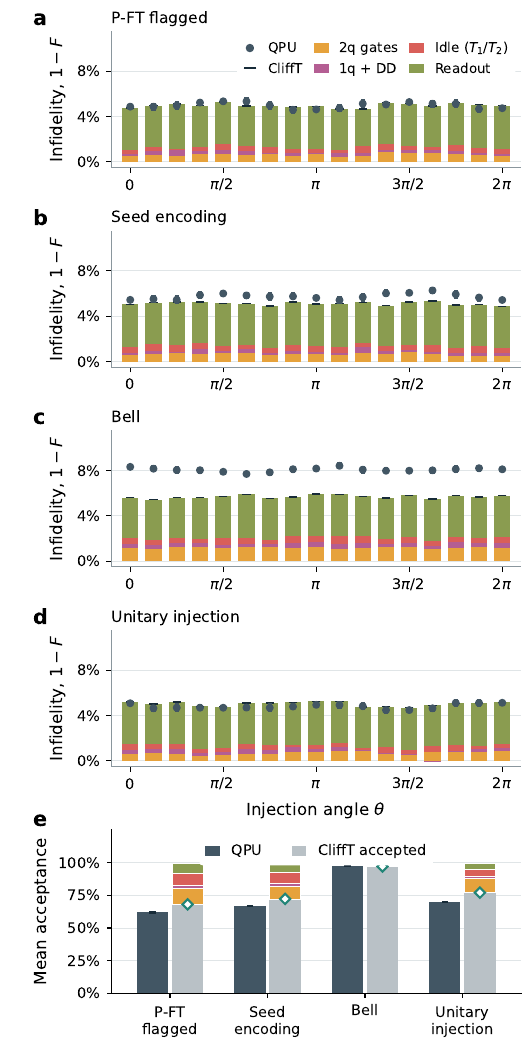}
\caption{Color code injection compared across angles. \textbf{(a)} Flagged P-FT. \textbf{(b)} Seed encoding. \textbf{(c)} Bell injection. \textbf{(d)} Unitary injection. IBM logical state infidelities are compared with the \texttt{CliffT} calibration model. Colored contributions attribute predicted errors to two qubit gates, single qubit gates including dynamical decoupling, idle noise, and readout. \textbf{(e)} Mean acceptance over the 17 angles, with gray model acceptance and colored contributions to rejection. Error bars show $95\%$ statistical intervals. Complete injection circuits are linked in Crumble (\protect\href{https://algassert.com/crumble\#circuit=Q(6,2)26;Q(7,2)27;Q(6,3)36;Q(7,3)37;Q(8,3)38;Q(9,3)39;Q(6,4)46;Q(7,4)47;Q(8,4)48;Q(9,4)49;Q(6,5)56;Q(7,5)57;Q(8,5)58;Q(6,6)66;R\_47;S\_47;SQRT\_X\_47;S\_47;TICK;R\_36;TICK;R\_57;S\_57;SQRT\_X\_57;S\_57;TICK;R\_66;TICK;R\_38;TICK;R\_26;TICK;R\_49;S\_49;SQRT\_X\_49;S\_49;TICK;R\_46;TICK;S\_46;SQRT\_X\_46;S\_46;CZ\_47\_46;S\_46;SQRT\_X\_46;S\_46\_36;SQRT\_X\_36;S\_36;CZ\_46\_36;S\_46;SQRT\_X\_46;S\_46;CZ\_47\_46;S\_46;SQRT\_X\_46;S\_46;R\_46;S\_46;SQRT\_X\_46;S\_46;TICK;R\_56;TICK;S\_56;SQRT\_X\_56;S\_56;CZ\_57\_56;S\_56;SQRT\_X\_56;S\_56\_66;SQRT\_X\_66;S\_66;CZ\_56\_66;S\_56;SQRT\_X\_56;S\_56;CZ\_57\_56;TICK;SQRT\_ZZ\_47\_57;TICK;S\_56;SQRT\_X\_56;S\_56;R\_56;TICK;S\_56;SQRT\_X\_56;S\_56;CZ\_46\_56;S\_47;SQRT\_X\_47;S\_47;CZ\_46\_47\_46\_36;S\_56;SQRT\_X\_56;S\_56;CZ\_56\_66;S\_57;SQRT\_X\_57;S\_57;CZ\_56\_57;S\_66;SQRT\_X\_66;S\_66\_56;SQRT\_X\_56;S\_56;CZ\_66\_56;R\_48;S\_48;SQRT\_X\_48;S\_48;TICK;R\_58;TICK;S\_58;SQRT\_X\_58;S\_58;CZ\_48\_58;S\_58;SQRT\_X\_58;S\_58;CZ\_58\_57;S\_57;SQRT\_X\_57;S\_57\_58;SQRT\_X\_58;S\_58;CZ\_57\_58\_57\_56;R\_27;S\_27;SQRT\_X\_27;S\_27;TICK;R\_37;TICK;S\_37;SQRT\_X\_37;S\_37;CZ\_27\_37;S\_26;SQRT\_X\_26;S\_26;CZ\_27\_26;S\_26;SQRT\_X\_26;S\_26\_27;SQRT\_X\_27;S\_27;CZ\_26\_27;S\_37;SQRT\_X\_37;S\_37\_38;SQRT\_X\_38;S\_38;CZ\_37\_38\_48\_38\_37\_47\_48\_47;S\_47;SQRT\_X\_47;S\_47\_48;SQRT\_X\_48;S\_48;CZ\_47\_48;S\_38;SQRT\_X\_38;S\_38;CZ\_38\_48\_37\_36;S\_36;SQRT\_X\_36;S\_36\_37;SQRT\_X\_37;S\_37;CZ\_36\_37\_47\_37;S\_46;SQRT\_X\_46;S\_46;CZ\_36\_46\_47\_46\_38\_37;S\_46;SQRT\_X\_46;S\_46;CZ\_46\_56;S\_27;SQRT\_X\_27;S\_27;CZ\_27\_37;R\_39;S\_39;SQRT\_X\_39;S\_39;TICK;S\_49;SQRT\_X\_49;S\_49;CZ\_39\_49;S\_49;SQRT\_X\_49;S\_49;CZ\_49\_48;S\_48;SQRT\_X\_48;S\_48;CZ\_48\_58;S\_39;SQRT\_X\_39;S\_39\_46;SQRT\_X\_46;S\_46\_27;SQRT\_X\_27;S\_27\_48;SQRT\_X\_48;S\_48\_56;SQRT\_X\_56;S\_56\_47;SQRT\_X\_47;S\_47\_36;SQRT\_X\_36;S\_36\_66;SQRT\_X\_66;S\_66\_57;SQRT\_X\_57;S\_57\_37;SQRT\_X\_37;S\_37\_26;SQRT\_X\_26;S\_26\_38;SQRT\_X\_38;S\_38\_58;SQRT\_X\_58;S\_58\_49;SQRT\_X\_49;S\_49;TICK;M\_39\_46\_27\_48\_56\_37\_58;TICK;R\_46;S\_46;SQRT\_X\_46;S\_46;TICK;R\_27;S\_27;SQRT\_X\_27;S\_27;TICK;R\_48;S\_48;SQRT\_X\_48;S\_48;TICK;R\_56;TICK;S\_56;SQRT\_X\_56;S\_56;CZ\_46\_56\_46\_47\_46\_36;S\_56;SQRT\_X\_56;S\_56;CZ\_56\_66\_56\_57;S\_66;SQRT\_X\_66;S\_66\_56;SQRT\_X\_56;S\_56;CZ\_66\_56;R\_37;TICK;S\_37;SQRT\_X\_37;S\_37;CZ\_27\_37\_27\_26;S\_26;SQRT\_X\_26;S\_26\_27;SQRT\_X\_27;S\_27;CZ\_26\_27;S\_37;SQRT\_X\_37;S\_37;CZ\_37\_38\_37\_47\_37\_36;S\_36;SQRT\_X\_36;S\_36\_37;SQRT\_X\_37;S\_37;CZ\_36\_37;S\_46;SQRT\_X\_46;S\_46;CZ\_36\_46;R\_58;TICK;S\_58;SQRT\_X\_58;S\_58;CZ\_48\_58\_48\_49\_48\_38\_48\_47;S\_47;SQRT\_X\_47;S\_47\_48;SQRT\_X\_48;S\_48;CZ\_47\_48\_47\_37\_47\_46;S\_38;SQRT\_X\_38;S\_38;CZ\_38\_48\_38\_37;S\_49;SQRT\_X\_49;S\_49;CZ\_49\_48;S\_58;SQRT\_X\_58;S\_58;CZ\_58\_57;S\_57;SQRT\_X\_57;S\_57\_58;SQRT\_X\_58;S\_58;CZ\_57\_58\_57\_56;TICK;S\_46;SQRT\_X\_46;S\_46;CZ\_46\_56;S\_48;SQRT\_X\_48;S\_48;CZ\_48\_58;S\_27;SQRT\_X\_27;S\_27;CZ\_27\_37;S\_46;SQRT\_X\_46;S\_46\_27;SQRT\_X\_27;S\_27\_48;SQRT\_X\_48;S\_48\_56;SQRT\_X\_56;S\_56\_37;SQRT\_X\_37;S\_37\_58;SQRT\_X\_58;S\_58;TICK;M\_46\_27\_48\_56\_37\_58;TICK\_}{PQEC flagged}; \protect\href{https://algassert.com/crumble\#circuit=Q(6,2)26;Q(6,3)36;Q(8,3)38;Q(6,4)46;Q(7,4)47;Q(8,4)48;Q(9,4)49;Q(6,5)56;Q(7,5)57;Q(8,5)58;Q(9,5)59;Q(6,6)66;R\_66;S\_66;SQRT\_X\_66;S\_66;TICK;TICK;S\_DAG\_66;TICK;R\_56;S\_56;SQRT\_X\_56;S\_56;TICK;R\_57;TICK;R\_58;S\_58;SQRT\_X\_58;S\_58;TICK;R\_46;TICK;S\_46;SQRT\_X\_46;S\_46;CZ\_56\_46;S\_56;SQRT\_X\_56;S\_56;CZ\_66\_56;R\_47;TICK;R\_59;TICK;S\_59;SQRT\_X\_59;S\_59;CZ\_58\_59;S\_57;SQRT\_X\_57;S\_57;CZ\_58\_57;S\_57;SQRT\_X\_57;S\_57;CZ\_57\_56;S\_56;SQRT\_X\_56;S\_56\_66;SQRT\_X\_66;S\_66;CZ\_56\_66;S\_58;SQRT\_X\_58;S\_58;CZ\_57\_58;S\_57;SQRT\_X\_57;S\_57;CZ\_56\_57;R\_49;TICK;S\_59;SQRT\_X\_59;S\_59\_49;SQRT\_X\_49;S\_49;CZ\_59\_49;S\_49;SQRT\_X\_49;S\_49\_59;SQRT\_X\_59;S\_59;CZ\_49\_59;R\_36;S\_36;SQRT\_X\_36;S\_36;TICK;R\_48;S\_48;SQRT\_X\_48;S\_48;TICK;S\_47;SQRT\_X\_47;S\_47;CZ\_48\_47;R\_38;TICK;S\_38;SQRT\_X\_38;S\_38;CZ\_48\_38;S\_38;SQRT\_X\_38;S\_38\_48;SQRT\_X\_48;S\_48;CZ\_38\_48;R\_26;TICK;S\_26;SQRT\_X\_26;S\_26;CZ\_36\_26\_36\_46;S\_47;SQRT\_X\_47;S\_47;CZ\_47\_46;S\_46;SQRT\_X\_46;S\_46\_36;SQRT\_X\_36;S\_36;CZ\_46\_36;S\_47;SQRT\_X\_47;S\_47;CZ\_46\_47;S\_46;SQRT\_X\_46;S\_46;CZ\_56\_46;S\_66;SQRT\_X\_66;S\_57;SQRT\_X\_57;S\_47;SQRT\_X\_47;S\_DAG\_49;S\_36;SQRT\_X\_36;S\_DAG\_38;S\_26;SQRT\_X\_26;TICK;S\_58;SQRT\_X\_58;S\_58\_59;SQRT\_X\_59;S\_59\_48;SQRT\_X\_48;S\_48\_46;SQRT\_X\_46;S\_46\_56;SQRT\_X\_56;S\_56;TICK;M\_58\_59\_48\_46\_56;TICK\_}{Bell injection}; \protect\href{https://algassert.com/crumble\#circuit=Q(6,2)26;Q(7,2)27;Q(6,3)36;Q(7,3)37;Q(8,3)38;Q(6,4)46;Q(7,4)47;Q(8,4)48;Q(9,4)49;Q(6,5)56;Q(7,5)57;Q(8,5)58;Q(9,5)59;Q(6,6)66;R\_66;S\_66;SQRT\_X\_66;S\_66;TICK;R\_56;S\_56;SQRT\_X\_56;S\_56;TICK;R\_57;S\_57;SQRT\_X\_57;S\_57;TICK;R\_58;S\_58;SQRT\_X\_58;S\_58;TICK;R\_59;S\_59;SQRT\_X\_59;S\_59;TICK;R\_46;TICK;R\_47;S\_47;SQRT\_X\_47;S\_47;TICK;R\_48;TICK;R\_49;TICK;R\_36;S\_36;SQRT\_X\_36;S\_36;TICK;S\_46;SQRT\_X\_46;S\_46;CZ\_36\_46;R\_37;TICK;S\_37;SQRT\_X\_37;S\_37;CZ\_47\_37\_36\_37;R\_38;S\_38;SQRT\_X\_38;S\_38;TICK;S\_48;SQRT\_X\_48;S\_48;CZ\_38\_48\_47\_48\_58\_48\_47\_46\_56\_46;S\_46;SQRT\_X\_46;S\_46\_56;SQRT\_X\_56;S\_56;CZ\_46\_56;R\_46;TICK;S\_48;SQRT\_X\_48;S\_48\_58;SQRT\_X\_58;S\_58;CZ\_48\_58\_57\_58\_57\_56;S\_58;SQRT\_X\_58;S\_58\_57;SQRT\_X\_57;S\_57;CZ\_58\_57;S\_56;SQRT\_X\_56;S\_56;CZ\_56\_57;S\_57;SQRT\_X\_57;S\_57;TICK;S\_DAG\_57;TICK;S\_57;SQRT\_X\_57;S\_57;CZ\_56\_57;S\_56;SQRT\_X\_56;S\_56;CZ\_66\_56;S\_56;SQRT\_X\_56;S\_56\_66;SQRT\_X\_66;S\_66;CZ\_56\_66;R\_56;S\_56;SQRT\_X\_56;S\_56;TICK;S\_46;SQRT\_X\_46;S\_46;CZ\_56\_46\_56\_66\_58\_57\_56\_57;S\_66;SQRT\_X\_66;S\_66\_56;SQRT\_X\_56;S\_56;CZ\_66\_56;S\_58;SQRT\_X\_58;S\_58;CZ\_59\_58;S\_58;SQRT\_X\_58;S\_58\_59;SQRT\_X\_59;S\_59;CZ\_58\_59;R\_58;S\_58;SQRT\_X\_58;S\_58;TICK;S\_59;SQRT\_X\_59;S\_59\_49;SQRT\_X\_49;S\_49;CZ\_59\_49;S\_46;SQRT\_X\_46;S\_46\_36;SQRT\_X\_36;S\_36;CZ\_46\_36;R\_48;TICK;S\_48;SQRT\_X\_48;S\_48;CZ\_58\_48\_58\_57;S\_57;SQRT\_X\_57;S\_57\_58;SQRT\_X\_58;S\_58;CZ\_57\_58\_57\_56;S\_49;SQRT\_X\_49;S\_49\_59;SQRT\_X\_59;S\_59;CZ\_49\_59\_38\_37;S\_48;SQRT\_X\_48;S\_48\_38;SQRT\_X\_38;S\_38;CZ\_48\_38;S\_49;SQRT\_X\_49;S\_49;CZ\_48\_49;R\_26;S\_26;SQRT\_X\_26;S\_26;TICK;R\_27;S\_27;SQRT\_X\_27;S\_27;TICK;CZ\_27\_37;S\_37;SQRT\_X\_37;S\_37\_27;SQRT\_X\_27;S\_27;CZ\_37\_27;R\_37;S\_37;SQRT\_X\_37;S\_37;TICK;CZ\_26\_27;S\_27;SQRT\_X\_27;S\_27\_26;SQRT\_X\_26;S\_26;CZ\_27\_26;R\_27;TICK;S\_27;SQRT\_X\_27;S\_27;CZ\_37\_27;S\_47;SQRT\_X\_47;S\_47;CZ\_37\_47\_46\_47\_48\_47;S\_47;SQRT\_X\_47;S\_47\_48;SQRT\_X\_48;S\_48;CZ\_47\_48\_37\_38\_37\_36;S\_36;SQRT\_X\_36;S\_36\_37;SQRT\_X\_37;S\_37;CZ\_36\_37\_47\_37;S\_46;SQRT\_X\_46;S\_46;CZ\_36\_46\_47\_46;S\_56;SQRT\_X\_56;S\_56;CZ\_56\_46;S\_38;SQRT\_X\_38;S\_38;CZ\_38\_48;S\_49;SQRT\_X\_49;S\_49;CZ\_49\_48;S\_58;SQRT\_X\_58;S\_58;CZ\_58\_48\_38\_37;S\_27;SQRT\_X\_27;S\_27;CZ\_27\_26;S\_26;SQRT\_X\_26;S\_26\_27;SQRT\_X\_27;S\_27;CZ\_26\_27;S\_37;SQRT\_X\_37;S\_37;CZ\_37\_27;S\_DAG\_66\_57\_47\_49\_36\_38\_26;TICK;S\_46;SQRT\_X\_46;S\_46\_27;SQRT\_X\_27;S\_27\_48;SQRT\_X\_48;S\_48\_59;SQRT\_X\_59;S\_59\_56;SQRT\_X\_56;S\_56\_37;SQRT\_X\_37;S\_37\_58;SQRT\_X\_58;S\_58;TICK;M\_46\_27\_48\_59\_56\_37\_58;TICK\_}{Unitary injection}; \protect\href{https://algassert.com/crumble\#circuit=Q(6,2)26;Q(6,3)36;Q(7,3)37;Q(8,3)38;Q(6,4)46;Q(7,4)47;Q(8,4)48;Q(9,4)49;Q(6,5)56;Q(7,5)57;Q(6,6)66;R\_26\_36\_37\_38\_46\_47\_48\_49\_56\_57\_66;H\_36\_38\_57\_47;S\_47;TICK;CX\_36\_46\_47\_46\_46\_56\_56\_46\_36\_26\_56\_66\_66\_56\_38\_48\_47\_48\_48\_49\_49\_48\_57\_47\_47\_37\_37\_47\_37\_36\_36\_26\_37\_38\_36\_37\_38\_37\_37\_47\_47\_37\_47\_57;M\_37\_46\_48\_56\_}{Seed encoding}).}
\label{fig:injection}
\end{figure}

Cultivation followed by kickback tomography gives a probability of $0.999859\%$ for the expected analyzer outcome, with $14.21\%$ acceptance. We then teleport a cultivated $T$ state onto a logical $|+\rangle$ input and apply the required $S$ and $Z$ corrections in real time. The final kickback measurement gives a conditional error probability of $1.93\%$. Combining two such teleportations with an initially cultivated $T$ state and transversal Hadamards produces the $THTHT$ state, with measured fidelities of $91.33\%$ for corrections in postprocessing and $90.98\%$ for application of corrections in real time. We also perform logical process tomography using six cultivated Pauli eigenstates. 

\subsection{Circuit Optimizations}

We choose the physical layout to connect resource preparation directly to lattice surgery. The surgery construction follows the color code merge and split protocol of Lacroix et al.~\cite{lacroix2025scaling}. We exchange the $X$ and $Z$ bases to measure $Z_LZ_L$ for resource state teleportation and vary the number of surgery rounds. The corresponding preparation and readout bases are exchanged consistently. This preserves the merge and split mechanism while providing the parity needed for the conditional $S$ correction.

The cultivation demonstration of Rosenfeld et al.~\cite{rosenfeld2026cultivation} uses a different assignment of data and helper sites. We adapt injection and cultivation to a common layout compatible with native $R_{ZZ}$ injection and the surgery interface. The seven data qubits then remain on the same physical sites through preparation, cultivation and surgery, avoiding an intermediate state transfer network. We resynthesize the seed encoding circuit and retain the fold and unfold structure of cultivation, adapting its interaction path and helper arrangement to this layout. The adapted network uses 22 entanglers and four helpers, compared with 24 entanglers and five helpers in the reference construction. We verify the ideal accepted measurement branches for both signs. The supplement presents the layout comparison and fault analysis of these adaptations.

Within this common layout, we reduce the cost of error detection by retaining native $R_{ZZ}$ injection gates and using fused measurement and reset for ancilla reuse~\cite{kan2026}. Adjacent single-qubit gates are combined during physical compilation using the Qiskit compiler \cite{qiskit2024}. We retain barriers only where needed to preserve the intended preparation sequence, since additional barriers restrict gate cancellation and scheduling. Preparation Hadamards are delayed until the first interaction so that qubits remain in $|0\rangle$ during preceding idle periods. We use fast readout, overlap compatible final data and syndrome measurements, and apply $XX$ dynamical decoupling during flag and syndrome measurement waits. We also do cultivation of the basis states during process tomography to prevent input basis bias.

We limit syndrome extraction to code projection, lattice surgery, and final readout. The additional operations in longer sequences can offset the benefit of repeated error detection at these distances. We compare one and three surgery rounds to assess this tradeoff with decoding. Simulations in \texttt{CliffT} without noise verify the logical output and deterministic measurement outcomes after compilation~\cite{chase_clifft_2026}. 

\subsection{Error Budgeting}

\begin{figure*}[!tp]
\centering
\begin{tikzpicture}[
  every node/.style={inner sep=0, outer sep=0}
]
  \node (a) {%
    \includegraphics[width=0.72\textwidth]
      {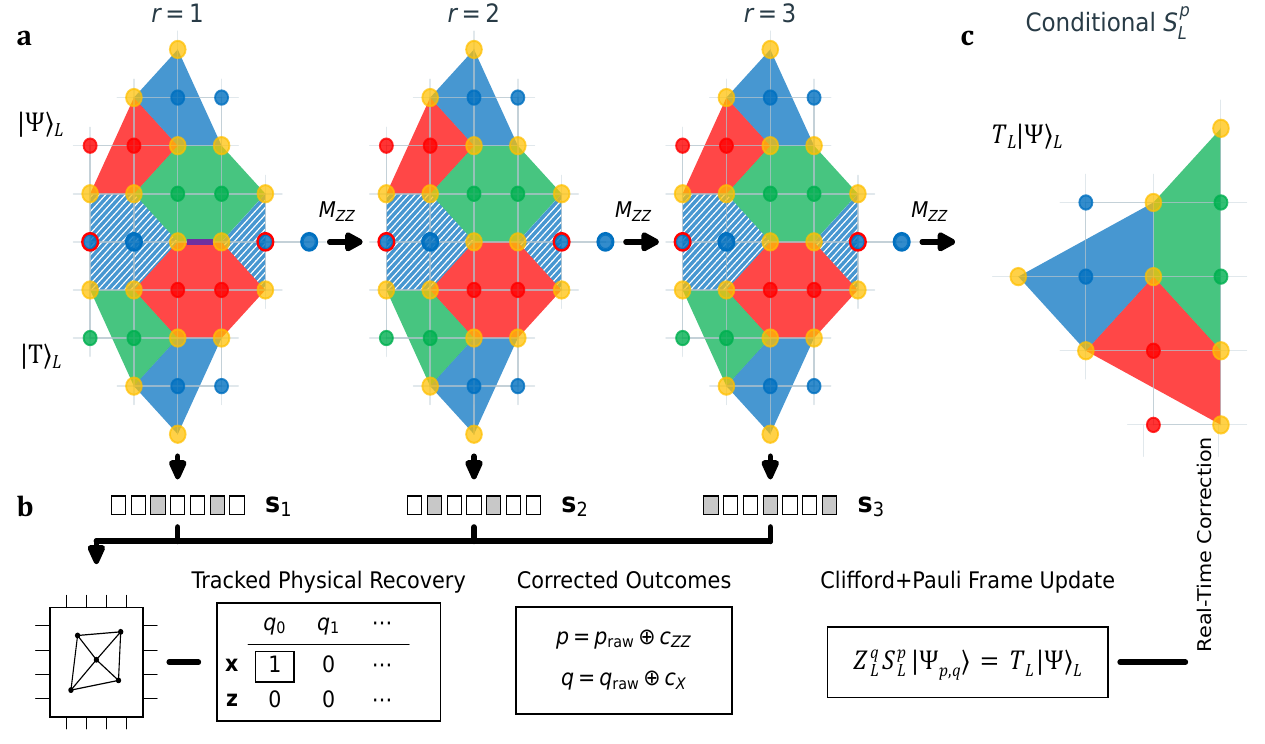}%
  };

  \node[right=0.005\textwidth of a] (b) {%
    \includegraphics[width=0.28\textwidth]
      {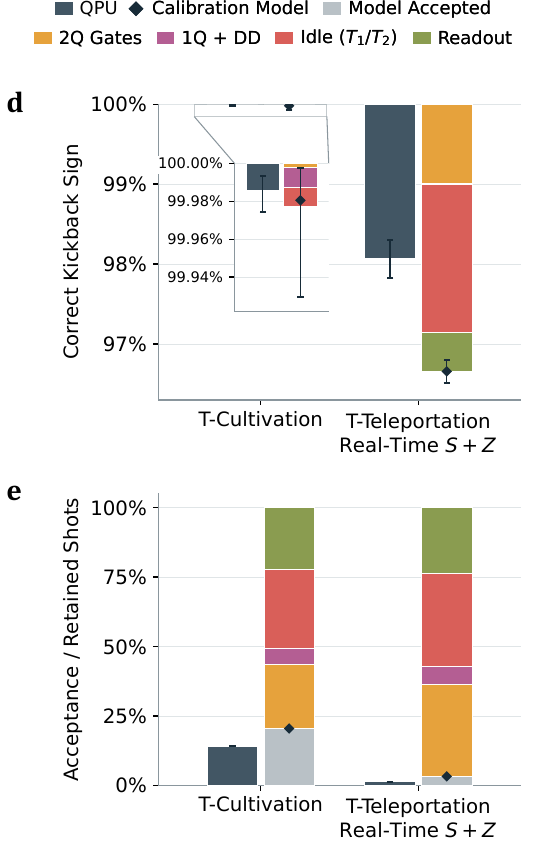}%
  };
\end{tikzpicture}
\caption{Consumption of a cultivated $T$ resource through color code teleportation with one surgery round and real-time correction. \textbf{(a)} Syndrome extraction across the lattice surgery seam. Blue rings identify Z check readouts, red rings identify flags, and purple links denote Bell pairs. \textbf{(b)} The decoder tracks a physical Pauli recovery and corrects the measured parities according to $p=p_{\rm raw}\oplus c_{ZZ}$ and $q=q_{\rm raw}\oplus c_X$. Here $p_{\rm raw}$ and $q_{\rm raw}$ are the measured logical $ZZ$ and resource $X$ outcomes; $c_{ZZ}$ and $c_X$ are the parity flips inferred from the physical recovery. The corrected outcomes determine the logical corrections. \textbf{(c)} Conditional transversal $S$ and $Z$ gates produce $T_L|\psi_L\rangle$. \textbf{(d)} Kickback measurements compare the resource after cultivation with the output after teleportation and real-time correction, testing whether its quality is retained when consumed. The measured probabilities are $99.9859\%$ ($95\%$ CI: $99.9741\%$ to $99.9933\%$) and $98.0693\%$ ($95\%$ CI: $97.8210\%$ to $98.2962\%$), respectively. \textbf{(e)} Corresponding acceptance probabilities of $14.212\%$ and $1.3311\%$. IBM measurements are compared with the \texttt{CliffT} model constructed from calibration data. Colored contributions attribute the predicted error and rejection to two qubit gates, single qubit gates including dynamical decoupling, idle noise, and readout. Gray denotes predicted acceptance. IBM error bars show $95\%$ binomial confidence intervals. Model error bars show $95\%$ sampling intervals; colored attribution segments are shown without individual intervals. The Crumble circuit for lattice surgery is \protect\href{https://algassert.com/crumble\#circuit=Q(5,6)65;Q(4,7)74;Q(5,7)75;Q(6,7)76;Q(2,8)82;Q(3,8)83;Q(4,8)84;Q(5,8)85;Q(6,8)86;Q(7,8)87;Q(8,8)88;Q(1,9)91;Q(2,9)92;Q(3,9)93;Q(4,9)94;Q(5,9)95;Q(6,9)96;Q(7,9)97;Q(8,9)98;Q(9,9)99;Q(2,10)102;Q(3,10)103;Q(4,10)104;Q(5,10)105;Q(6,10)106;Q(7,10)107;Q(8,10)108;Q(3,11)113;Q(4,11)114;Q(5,11)115;Q(6,11)116;Q(7,11)117;R\_65\_75\_82\_84\_85\_86\_88\_92\_94\_95\_96\_98\_103\_105\_107\_113\_115\_117;H\_86\_88\_107\_103\_82\_84;R\_85\_95;H\_85;CX\_85\_95;TICK;CX\_86\_96\_88\_98\_107\_117\_103\_113\_82\_92\_84\_94\_86\_76\_96\_106\_98\_108\_107\_97\_103\_93\_92\_102\_84\_74\_94\_104\_86\_87\_96\_97\_98\_99\_107\_108\_103\_102\_92\_91\_84\_83\_94\_93\_86\_85\_96\_95\_88\_87\_98\_97\_107\_106\_117\_116\_103\_104\_113\_114\_82\_83\_92\_93\_84\_85\_94\_95\_85\_84\_95\_94\_85\_86\_95\_96\_87\_88\_97\_98\_106\_107\_116\_117\_104\_103\_114\_113\_83\_82\_93\_92\_76\_86\_106\_96\_108\_98\_97\_107\_93\_103\_102\_92\_74\_84\_104\_94\_87\_86\_97\_96\_99\_98\_108\_107\_102\_103\_91\_92\_83\_84\_93\_94\_86\_96\_88\_98\_107\_117\_103\_113\_82\_92\_84\_94;H\_86\_88\_107\_103\_82\_84;TICK;M\_96;R\_96;M\_98;R\_98;M\_117;R\_117;M\_113;R\_113;M\_92;R\_92;M\_94;R\_94;M\_86;R\_86;M\_88;R\_88;M\_107;R\_107;M\_103;R\_103;M\_82;R\_82;M\_84;R\_84;TICK;H\_86\_88\_107\_103\_82\_84;R\_65;H\_65;R\_105;H\_105;R\_75\_115;TICK;CX\_65\_75\_105\_115\_76\_75\_106\_105\_116\_115\_74\_75\_104\_105\_114\_115\_85\_75\_95\_105\_65\_75\_105\_115;H\_65\_105;TICK;M\_75;R\_75;M\_115;R\_115;M\_65;R\_65;M\_105;R\_105;TICK;H\_65\_105;TICK;CX\_85\_95;H\_85;TICK;M\_85;R\_85;M\_95;R\_95;TICK;TICK;H\_74\_83\_93\_91\_104\_102\_114;TICK;M\_74\_83\_93\_91\_104\_102\_114\_}{here}. The Crumble circuit for cultivation is \protect\href{https://algassert.com/crumble\#circuit=Q(6,2)26;Q(7,2)27;Q(6,3)36;Q(7,3)37;Q(8,3)38;Q(7,4)47;Q(8,4)48;Q(9,4)49;Q(6,5)56;Q(7,5)57;Q(6,6)66;TICK;R\_48\_56\_37\_27;S\_DAG\_47\_36\_57\_66\_38\_26\_49;S\_DAG\_47\_36\_57\_66\_38\_26\_49;H\_47\_36\_57\_66\_38\_26\_49\_27;TICK;CX\_27\_37\_26\_36\_36\_37\_38\_37\_66\_56\_56\_57\_49\_48\_48\_47\_57\_47\_47\_37\_27\_37;H\_27;X\_37;M\_37\_27;R\_37\_27;X\_37;H\_27;TICK;CX\_27\_37\_47\_37\_57\_47\_48\_47\_49\_48\_56\_57\_66\_56\_38\_37\_36\_37\_26\_36\_27\_37;H\_27\_47\_36\_57\_66\_38\_26\_49;S\_47\_36\_57\_66\_38\_26\_49;S\_47\_36\_57\_66\_38\_26\_49;M\_48\_56\_37\_27;TICK\_}{here}.}
\label{fig:real_time}
\end{figure*}

With the circuits compiled, we compare their measured infidelity and acceptance with a stochastic Pauli model to assess which errors the calibration data can explain. Such models are widely used for injection and cultivation, but omit effects such as coherent errors and leakage~\cite{hines2026simulating}. We simulate the compiled protocols with \texttt{CliffT}, using saved IBM calibration data rather than fitting measured fidelities. We include depolarizing gate errors, Pauli channels for idle noise obtained by twirling relaxation and dephasing, and classical readout assignment errors. Simulation and experiment use matched deterministic checks and frame interpretation.

We construct the error budget using Shapley attribution~\cite{shapley1953value}. Define $\mathcal G$ for two qubit errors, single qubit errors including dynamical decoupling, idle noise, and readout errors. We first simulate all 16 subsets $S\subseteq\mathcal G$, enabling only the noise sources in $S$. The compiled gates, timing and analysis rules remain fixed, with ideal reset.

For each simulation, we apply the prescribed postselection and frame corrections. Let $A(S)$ be the accepted fraction and $F(S)$ the fidelity reconstructed from accepted outcomes. We calculate the loss as $v(S)=1-F(S)$ for infidelity or $v(S)=1-A(S)$ for acceptance loss. For kickback error budgets, $v(S)$ is the conditional wrong sign probability.

We then calculate the change in loss when source $i\in\mathcal G$ is added to each subset that excludes it. Averaging these changes gives its contribution,
\begin{equation}
\phi_i[v]=\sum_{S\subseteq\mathcal G\setminus\{i\}}
\frac{|S|!\,(3-|S|)!}{4!}
\bigl[v(S\cup\{i\})-v(S)\bigr].
\label{eq:shapley_budget}
\end{equation}
Here $|S|$ counts the sources already enabled, and the factorial weight averages over all 24 source orders. This distributes the effects of interactions among the four sources.

The contributions sum to $v(\mathcal G)-v(\varnothing)$, where $v(\varnothing)$ is the noiseless baseline. Adding this baseline recovers the full model prediction and keeps intrinsic rejection separate from rejection caused by noise. We obtain statistical intervals by bootstrap resampling the outcome counts and repeating the loss calculation and attribution. Contributions can be negative because enabling noise changes which shots survive postselection.

\subsection{Injection}

We first examine injection, which sets the initial error and acceptance for both partial QEC and cultivation. For direct analog rotations, the protocol must prepare useful states over a range of angles. We compare direct and flagged Partial-FT injection, the injection circuit used in Rosenfeld et. al. in the $XY$ plane, Bell injection~\cite{jones2016gauge}, and unitary injection~\cite{itogawa2025efficient}. We sweep $0\leq\theta\leq2\pi$ in steps of $\pi/8$, with $100{,}000$ shots per contributing Pauli basis. For states in either plane, the two corresponding logical Pauli expectation values determine the target-state fidelity. All protocols use data qubits $\{26,36,38,47,49,57,66\}$ on \processor~and retain shots passing their prescribed deterministic checks and flags.

Fig.~\ref{fig:injection} shows the tradeoff between infidelity and acceptance. The mean infidelities are $4.98\%$ for flagged Partial-FT, $5.74\%$  for Rosenfeld \textit{et al.} adapted $XY$ injection, $8.07\%$ for Bell injection, and $4.80\%$ for unitary injection. Corresponding acceptance probabilities are $62.0\%$, $67.1\%$,  $97.7\%$, and $69.7\%$. Flagged Partial-FT reduces infidelity relative to direct Partial-FT at lower acceptance, while unitary injection has the lowest measured mean infidelity. Figure~\ref{fig:injection} also shows less variation with angle for flagged Partial-FT: the measured infidelity ranges from $4.59\%$ to $5.34\%$, compared with $5.14\%$ to $7.25\%$ for direct Partial-FT. Bell injection illustrates a different tradeoff: its high acceptance is accompanied by the largest mean infidelity. 

We compare these results with the Pauli model constructed from the calibration data in Fig.~\ref{fig:injection}. Predicted mean infidelities are $4.97\%$ for flagged Partial-FT and $5.02\%$ for unitary injection, close to experiment. Bell injection shows the largest discrepancy, with $5.69\%$ predicted versus $8.07\%$ measured. In the model, readout contributes $3.64$--$3.71$ percentage points and is the largest contribution for every protocol. Logical readout therefore accounts for much of the predicted infidelity. Even where infidelity agrees, acceptance can differ: flagged Partial-FT has $68.2\%$ predicted acceptance versus $62.0\%$ measured.  The acceptance comparison tests whether it also predicts how often the preparation checks reject an attempt.

\begin{figure*}[!tp]
\centering
\begin{tikzpicture}[
  every node/.style={inner sep=0, outer sep=0}
]
  \node (a) {%
    \includegraphics[width=0.37\textwidth]
      {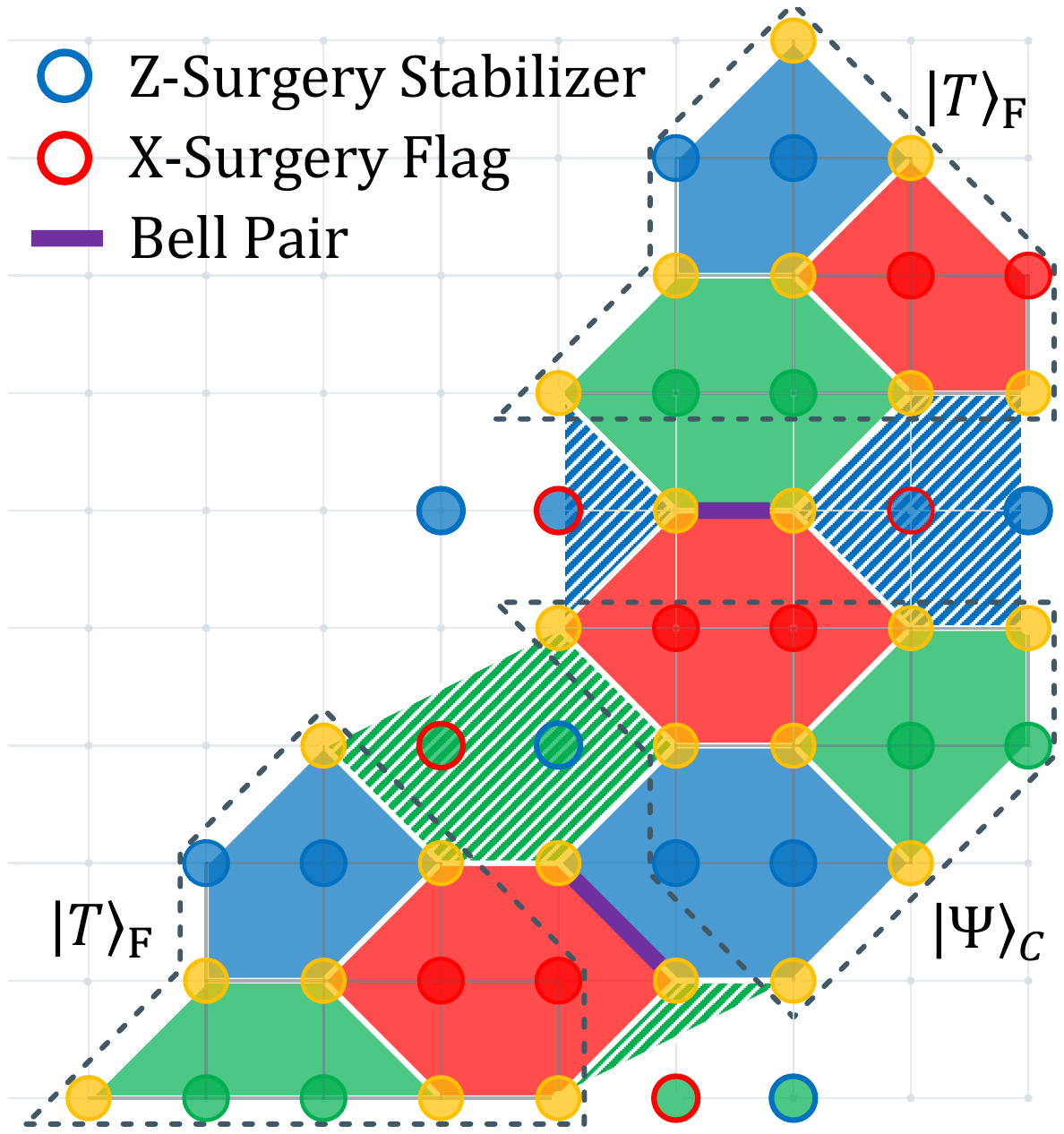}%
  };

  \node[right=0.005\textwidth of a] (b) {%
    \includegraphics[width=0.37\textwidth]
      {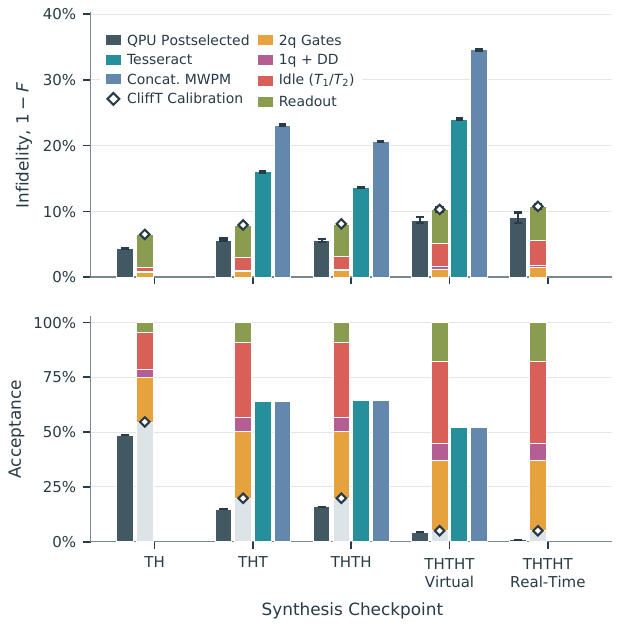}%
  };

  \node[right=0.005\textwidth of b] (c) {%
    \includegraphics[width=0.22\textwidth]
      {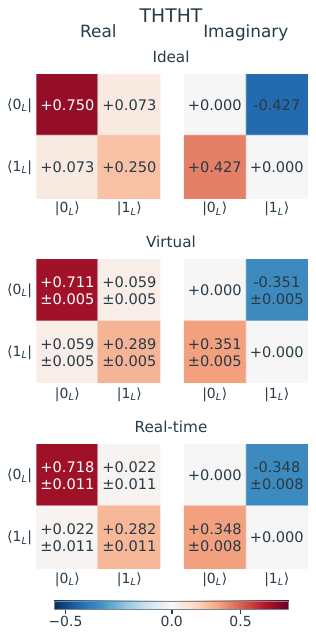}%
  };

  \node[fit=(a)(b)(c)] (panels) {};

    \foreach \p in {a,b,c} {
      \node[
        anchor=south west,
        inner sep=0,
        outer sep=0,
        yshift=-24pt,
        font=\bfseries
      ] at (\p.west |- panels.north) {\p};
    }
\end{tikzpicture}

\caption{\textbf{(a)} Logical synthesis implementation using two factories and one compute patch. $T_L$ gates are implemented via $\ket{T}$ teleportation onto a cultivated $\ket{T}_L$ state, with transversal $H^{\otimes7}$ on the compute patch between teleportations to synthesize nontrivial magic states. \textbf{(b)} Fidelity and acceptance of synthesis states at each successive checkpoint compared to calibration predictions. \textbf{(c)} Final density matrices comparing the ideal synthesized state versus the states corrected with postprocessing and real-time feed-forward $S_L$ respectively.}
\label{fig:preparation}
\end{figure*}

\subsection{Cultivation and Kickback Tomography}

The angle sweep establishes the performance of injection alone. For the fixed $T$ state, cultivation adds a logical check that can reject errors remaining after injection. We apply the fold and unfold circuits of Gidney et al.~\cite{gidney2024cultivation} and characterize the accepted states using logical state tomography and kickback tomography~\cite{rosenfeld2026cultivation}. For $XY$ states, the logical check distinguishes eigenstates of $M_T=(X_L+Y_L)/\sqrt2$.

The Rosenfeld \textit{et al.}~\cite{rosenfeld2026cultivation} cultivation demonstration circuit uses a slightly modified layout of the color code based off of the placement of their readout line to optimize the cultivation and kickback circuits. In order to use color code lattice surgery circuits for our synthesis, we use the lattice surgery-compatible layout and adapt our cultivation circuits. We find that our kickback tomography circuits only support a distance 2 instead of distance 3. However simulations support that wrong sign probability is still a lower bound of the fidelity under the noise model as discussed in Appendix \ref{sec:kickback_construction}.

With $XY$ plane injection using the adapted Rosenfeld \textit{et al.} circuit, we observe 10 negative analyzer outcomes among $71{,}060$ accepted shots from $500{,}000$ attempts. The conditional kickback error probability is $1.41\times10^{-4}$, with a $95\%$ interval of $[6.75\times10^{-5},2.59\times10^{-4}]$, at $14.21\%$ acceptance.

Logical state tomography gives infidelities after cultivation of $4.45\%$--$5.02\%$ across injections. Kickback tomography includes additional analyzer postselection. The injection protocol with the lowest infidelity does not necessarily give the lowest error after cultivation. Choosing a factory input therefore requires measuring the complete preparation sequence, including its acceptance, rather than ranking injection circuits alone. An injection check can reject faults that cultivation would also detect, as well as faults that do not produce a logical error. Additional injection checks therefore need not improve the final conditional fidelity, although the measurements do not identify which faults change the ranking.

\subsubsection{Real Time Correction and Synthesis}

Having characterized cultivation, we test whether the resource can be consumed to apply a gate to a separate logical qubit. We use lattice surgery to teleport the $T$ operation onto the data. The joint logical $ZZ$ outcome determines a logical $S$ correction, and the $X$ measurement outcome on the resource patch determines a Pauli-$Z$ correction. The ongoing Pauli frame determines how the measurement outcomes are interpreted. The resulting teleportation byproducts determine the logical corrections, as illustrated in Fig.~\ref{fig:real_time}. We implement feedforward after a surgery round using QASM \texttt{if} statements and transversal physical gates. The hardware feedforward is conditioned on measured parities. The $Z$ byproduct is accounted for by Pauli frame tracking~\cite{paler_software_2014}. A pending $S$ correction can instead change the required measurement basis, since $S_LX_LS_L^\dagger=Y_L$. Applying the correction before the next operation avoids carrying this additional Clifford frame through the synthesis circuit. Decoding is performed offline. For a $|+\rangle_L$ input, teleportation with correction in real time gives a conditional kickback error probability of $1.93\%$ at $1.33\%$ acceptance. This result includes state preparation and the analyzer.

This conditional teleportation provides the next step toward synthesis: combining several resource states with transversal Clifford gates. We cultivate a $T$ state in the data patch and teleport two further cultivated resources, with transversal Hadamard operations between $T$ operations. Preparation of the second resource is scheduled to finish near its teleportation, reducing the idle time of the cultivated state. One surgery round per teleportation produces $T_LH_LT_LH_LT_L|+\rangle_L$. The $TH$, $THT$, $THTH$, and $THTHT$ checkpoints distinguish the state after a transversal Hadamard from the state after a resource is consumed. The checkpoints thus probe error accumulation through synthesis. Tomography gives final state fidelities of $91.33\%$ with corrections applied in postprocessing and $90.98\%$ with corrections applied in real time. The density matrices in Fig.~\ref{fig:preparation} retain the expected populations and imaginary coherence, with reduced coherence relative to the ideal state. Both correction schemes therefore preserve the main features of the synthesized state. Their fidelity intervals overlap, although they select different fractions of the acquired shots. Acceptance is $4.28\%$ and $0.93\%$, respectively. For corrections applied in postprocessing, the model predicts $10.30\%$ infidelity and $4.99\%$ acceptance, compared with $8.67\%$ and $4.28\%$ measured.

\subsection{Logical Process Tomography}

\begin{figure}[!tbp]
\centering
\includegraphics[width=\linewidth,height=0.54\textheight,keepaspectratio]{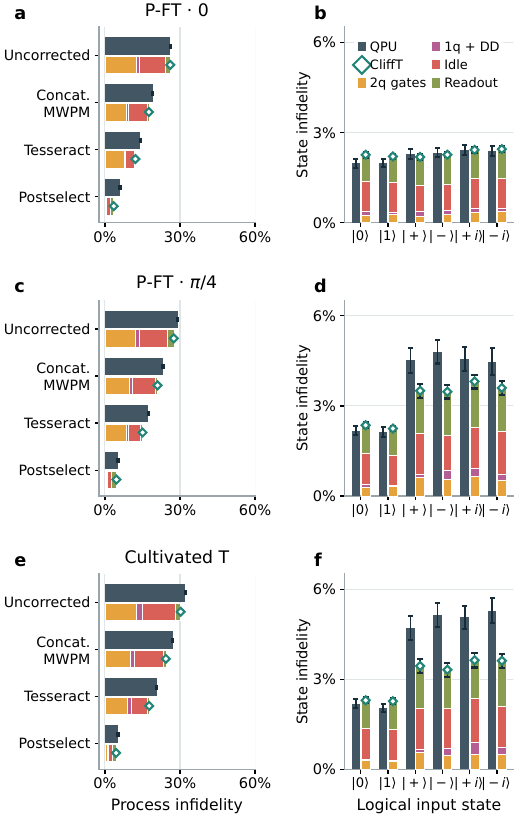}
\caption{One-round logical teleportation characterized by process and input state tomography. \textbf{(a)} Process infidelity for Partial-FT at $\theta=0$, comparing uncorrected reconstruction, Concatenated-MWPM, Tesseract and strict postselection. \textbf{(b)} Corresponding output state infidelities for the six Pauli eigenstate inputs under strict postselection. \textbf{(c,d)} The same comparisons for Partial-FT at $\theta=\pi/4$. \textbf{(e,f)} The same comparisons for a cultivated $T$ resource. All corrections are applied in postprocessing. \processor{} results are compared with calibrated \texttt{CliffT} simulations, with colored contributions from two qubit gates, single qubit gates including dynamical decoupling, idle noise and readout. Process results use the held-out test subset; input state results use the full acquisition. Error bars show approximate $95\%$ shot intervals for process infidelity and $95\%$ bootstrap intervals for input state infidelity. }
\label{fig:process_tomography}
\end{figure}

The synthesis experiment tests one sequence and one initial state. To characterize teleportation beyond that example, we measure its action on a complete set of logical inputs. We perform process tomography using all six logical Pauli eigenstates, prepared by cultivation, and measure the teleported outputs in $X_L$, $Y_L$, and $Z_L$. We test Partial-FT resources at $\theta=0$ and $\pi/4$ and cultivated $T$ resources, with one or three surgery rounds and virtual corrections. The reconstructed process includes preparation and measurement errors. Using all six inputs tests the action on both populations and coherences. In the input state panels of Fig.~\ref{fig:process_tomography}, Partial-FT at $\theta=0$ gives approximately $2\%$ output infidelity for every input. At $\theta=\pi/4$, equatorial inputs give approximately $4.5\%$--$4.8\%$, compared with approximately $2\%$ for $|0\rangle_L$ and $|1\rangle_L$. The input dependence is therefore small for the zero-angle operation but remains visible for the injected rotation.

The error attribution also predicts a dependence on the logical input. For Partial-FT at $\theta=\pi/4$, the measured infidelities averaged over opposite inputs are $2.16\%$ for $|0\rangle_L$ and $|1\rangle_L$, $4.67\%$ for $|+\rangle_L$ and $|-\rangle_L$, and $4.52\%$ for $|+i\rangle_L$ and $|-i\rangle_L$. The corresponding calibration predictions are $2.30\%$, $3.49\%$ and $3.71\%$. In the model, readout contributes $0.91$ percentage points for the computational basis inputs, compared with $1.43$ and $1.49$ percentage points for the $X$ and $Y$ inputs. Idle noise contributes $1.02$, $1.27$ and $1.39$ percentage points, respectively. Gate errors also make larger contributions for the superposition inputs. The model therefore reproduces the ordering of the input sensitivities, while the experiment shows a larger separation.

Different logical inputs reveal different components of the error channel. A logical $Z_L$ error preserves the computational basis states up to a global phase but changes their superpositions, while logical $X_L$ errors interchange $|0\rangle_L$ and $|1\rangle_L$. The observed ordering reflects the relative contributions of these error components together with preparation, postselection and readout. As code distance increases, differences arising from correctable physical faults should decrease when preparation, surgery and logical readout preserve the corresponding fault protection. In this regime, suppression of the residual logical errors brings the outputs for all six inputs closer to their ideal targets.

Detector error models constructed from the calibration data for the Clifford surgery and readout are decoded with Tesseract~\cite{beni2025tesseract} and Concatenated-MWPM~\cite{lee2025color, lee_low-overhead_2025}. The first measurements of newly introduced stabilizers can have random outcomes and are not postselected to zero. For Partial-FT with one surgery round at $\theta=0$, process infidelity decreases from $26.1\%$ without recovery to $19.2\%$ with Concatenated-MWPM and $14.2\%$ with Tesseract, with $43.2\%$ of shots retained. Strict postselection gives $5.92\%$ at $13.4\%$ acceptance. Figure~\ref{fig:process_tomography} shows that decoding recovers shots otherwise rejected by error detection. Strict postselection gives lower conditional infidelity at lower acceptance. The decoder comparison tests how much of the rejected population can be recovered. For Tesseract, the calibration model predicts $12.2\%$ for one round and $15.1\%$ for three rounds, compared with $14.2\%$ and $24.1\%$ measured.

We normalize the reconstructed Choi matrix $J_{\rm rec}$ to unit trace and calculate its overlap with the ideal process,
\begin{equation}
\widetilde J=\frac{J_{\rm rec}}{\operatorname{Tr}J_{\rm rec}},\quad
F_{\rm proc}=\operatorname{Tr}(\widetilde J J_U),\quad
\epsilon_{\rm proc}=1-F_{\rm proc},
\end{equation}
where $J_U=|\Phi_U\rangle\langle\Phi_U|$, $|\Phi_U\rangle=(U\otimes I)|\Phi^+\rangle$, and $|\Phi^+\rangle=(|00\rangle+|11\rangle)/\sqrt{2}$. Here $U=R_Z(\theta)$ is the target logical operation. The trace of $J_{\rm rec}$ gives the mean acceptance probability before normalization. 

\section{Conclusion}

Supplying accurate non-Clifford operations at an affordable resource cost is a central challenge for useful error-corrected computation. This work implements two approaches to that challenge within a common color code architecture on IBM~\processor{}: direct analog injection for partially fault-tolerant rotations, and cultivated resource states consumed through lattice surgery for logical synthesis. Our STAR-type protocol combines native $R_{ZZ}$ interactions with error detection to achieve $P_L=p/15+O(p^2)$ under the uniform circuit-level depolarizing model. Its local construction reduces the cost of preparing analog resource states on square grid hardware, making the physical interaction itself part of the solution to logical gate overhead.

The factory experiments connect resource preparation to the operations it enables. A common layout keeps the data qubits stationary through injection, cultivation, and lattice surgery, while transversal Clifford gates and real-time $S_L$ and $Z_L$ corrections allow cultivated resources to be consumed during computation. Kickback measurements give correct sign probabilities of $99.9859\%$ after cultivation and $98.0693\%$ after teleportation onto a logical $|+\rangle_L$ input. Successive resource consumption and transversal Hadamard gates then produce $T_LH_LT_LH_LT_L|+\rangle_L$, with final state fidelities above $90\%$ for both virtual and real-time corrections. This sequence demonstrates how small color code factories can supply an ongoing logical computation and assemble a non-T logical resource beyond individual cultivated $T$ states.

Logical process tomography characterizes resource teleportation across a complete set of inputs, and we demonstrate that Concatenated-MWPM and Tesseract suppress errors in the teleported logical operation. Together, these experiments establish a hardware setting in which analog injection and cultivated resource synthesis can be evaluated through their full preparation and consumption sequences. Their relative costs and accuracy will determine how early QEC processors implement the non-Clifford operations required by quantum algorithms. The combination of local resource preparation, transversal Clifford operations, and conditional lattice surgery provides an experimental foundation for developing that computational capability with color codes.

\section*{Acknowledgments}

This material is based upon work supported by the U.S. Department of Energy, Office of Science, National Quantum Information Science Research Centers, Quantum Science Center (QSC), and PNNL’s Quantum Algorithms and Architecture for Domain Science (QuAADS) Laboratory Directed Research and Development (LDRD) Initiative. 
This research used resources of the Oak Ridge Leadership Computing Facility (OLCF), a DOE Office of Science User Facility supported under Contract No. DE-AC05-00OR22725, and the National Energy Research Scientific Computing Center (NERSC), a DOE facility operated at Lawrence Berkeley National Laboratory under Contract No. DE-AC02-05CH11231, using NERSC award DDR-ERCAP0038957.
The Pacific Northwest National Laboratory is operated by Battelle for the U.S. Department of Energy under Contract No. DE-AC05-76RL01830. 

\section*{Contributions}

SG developed the color code rotation protocols and demonstrated their performance. SG and AH developed the magic state protocols and experiments to be run on~\processor. AH implemented the protocols and optimized the experiments. AH, SG, and SK all contributed to the analyses. AL advised SG and AH. SS advised AH, SK, and the implementation. CL advised SG and the overall work. All authors contributed to the final manuscript.

\bibliography{partial_ft}

\clearpage
\appendix

\section{Data Collection}
\label{sec:data_collection}

We constructed and compiled the circuits with Qiskit and submitted individual jobs to \texttt{ibm\_phoenix} through IBM Quantum Runtime. Each experimental setting specifies the injection protocol and angle, tomography basis, input state, syndrome rounds and correction mode. Shots for a setting were partitioned across jobs and combined after acquisition. Table~\ref{tab:paper_acquisitions} summarizes the hardware datasets used in the figures by execution day in UTC. Circuit counts refer to distinct experimental settings, with repeated shot partitions counted as one circuit.

\begin{table}[!bp]
\centering
\footnotesize
\setlength{\tabcolsep}{2pt}
\resizebox{\columnwidth}{!}{%
\begin{tabular}{@{}lcrr@{}}
\hline
Dataset & Day & Circuits & Shots ($10^6$) \\
\hline
Injection: main protocols & Oct 21 & 136 & 13.60 \\
Additional injection comparisons & Oct 21 & 68 & 6.80 \\
Cultivation tomography & Oct 21 & 12 & 1.20 \\
Cultivation kickback & Oct 21 & 6 & 3.00 \\
Real-time teleportation kickback & Oct 23 & 1 & 1.00 \\
Process tomography, $r=1$ & Oct 23 & 54 & 13.50 \\
TH through THTHT synthesis & Oct 27 & 15 & 14.86149 \\
\hline
Total & & 292 & 53.96149 \\
\hline
\end{tabular}}
\caption{Hardware datasets used in the paper figures. Dates are UTC in 2026. Counts give distinct circuit settings and raw shots before selection, with shared records counted once. Injection and cultivation tomography use $100{,}000$ shots per circuit, cultivation kickback uses $500{,}000$, and one-round process tomography uses $250{,}000$. Real-time teleportation kickback uses one million shots. Synthesis uses one million returned shots per circuit except THT-$X$, with $861{,}490$.}
\label{tab:paper_acquisitions}
\end{table}

We collected integrated complex IQ samples and QPU classified bits together using the QuantumProgram measurement level \texttt{both}. Each intermediate and terminal measurement was assigned a distinct classical destination, preserving the correspondence between its I/Q sample, classified outcome and shot. These records include syndrome and flag measurements as well as final data readout. The plotted tomography and kickback results use the classified records with the prescribed check selection and frame corrections; the associated IQ samples support analysis of the readout distributions.

For each circuit setting, we pool the measurement records from its shot partitions before evaluating acceptance and logical observables. Selection follows the deterministic checks specified by that protocol. Initially random gauge outcomes are retained and enter the frame interpretation, while flags and consistency checks select the accepted shots. Teleportation retains both logical outcome branches and applies their prescribed corrections. The figures use the classified records throughout; IQ samples remain associated with the same shots for readout analysis.

\section{Additional Injection Comparisons}
\label{sec:supplement_injection}

\begin{figure}[!tbp]
\centering
\includegraphics[width=\linewidth]{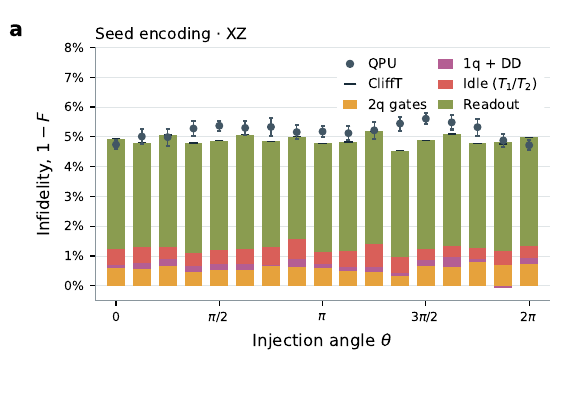}
\caption{Additional color code injection sweeps. Seed encoding in the $XZ$ plane. QPU state infidelities are compared with CliffT calibration predictions. Colored segments give the Shapley attribution to two qubit gates, single qubit gates including dynamical decoupling, idle noise and readout. QPU error bars show $95\%$ bootstrap intervals.}
\label{fig:supp_injection_controls}
\end{figure}

Figure~\ref{fig:supp_injection_controls} extends the injection sweep to seed encoding in the $XZ$ plane. This variant uses the same 17 angles from zero to $2\pi$ in steps of $\pi/8$, with $100{,}000$ shots per contributing Pauli basis. Together with the $XY$ seed encoding sweep in the main text, this comparison tests the choice of injection plane. The data qubit sites, calibration replay and prescribed selection follow the comparisons in the main text.

\section{Cultivation Comparison for Each Injection Protocol}
\label{sec:supplement_cultivation}

The injection sweep characterizes analog resource preparation over a range of angles. Here we fix the magic state angle at $\theta=\pi/4$ and compare the injection circuits as seeds for cultivation. The five variants are flagged PQEC, seed encoding in the $XY$ and $XZ$ planes, Bell injection and unitary injection. The two seed encoding variants correspond to the Google demonstration circuit adapted to the shared layout. Each seed is followed by the cultivation check for its target plane. This comparison tests which complete preparation sequence delivers the best accepted resource.

\begin{figure}[!tbp]
\centering
\includegraphics[width=\linewidth]{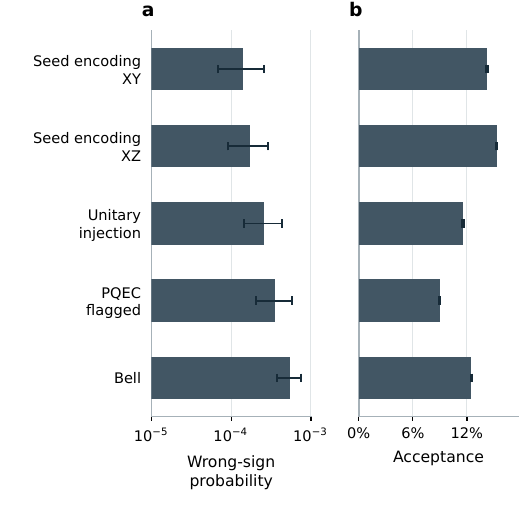}
\caption{Repeated kickback characterization of the cultivated resources. (a) Conditional negative sign probability for each injection seed. (b) Total acceptance through preparation and the analyzer checks. The $XY$ seed encoding circuit gives 10 negative outcomes among $71{,}060$ accepted shots. Error bars show exact $95\%$ binomial confidence intervals.}
\label{fig:supp_cultivation_kickback}
\end{figure}

We characterize all five cultivated outputs with repeated kickback measurements, using $500{,}000$ attempts per variant. Figure~\ref{fig:supp_cultivation_kickback} reports the negative sign probability among shots passing the analyzer checks, together with the fraction of attempts retained. This readout checks the logical magic state axis directly. Its two analyzers and their consistency checks define the final accepted ensemble.

The lowest measured kickback error probability occurs for $XY$ seed encoding, with 10 negative outcomes among $71{,}060$ accepted shots from $500{,}000$ attempts. The $XZ$ variant gives 13 negative outcomes among $76{,}438$ accepted shots, with overlapping confidence intervals. These outcomes motivate the $XY$ resource used in the subsequent consumption experiment. The circuit construction and simulations below describe how the analyzer checks contribute to interpreting this measurement.

\section{Cultivation and Kickback Circuit Construction}
\label{sec:kickback_construction}

\begin{figure}[!tbp]
\centering
\includegraphics[width=1\linewidth]{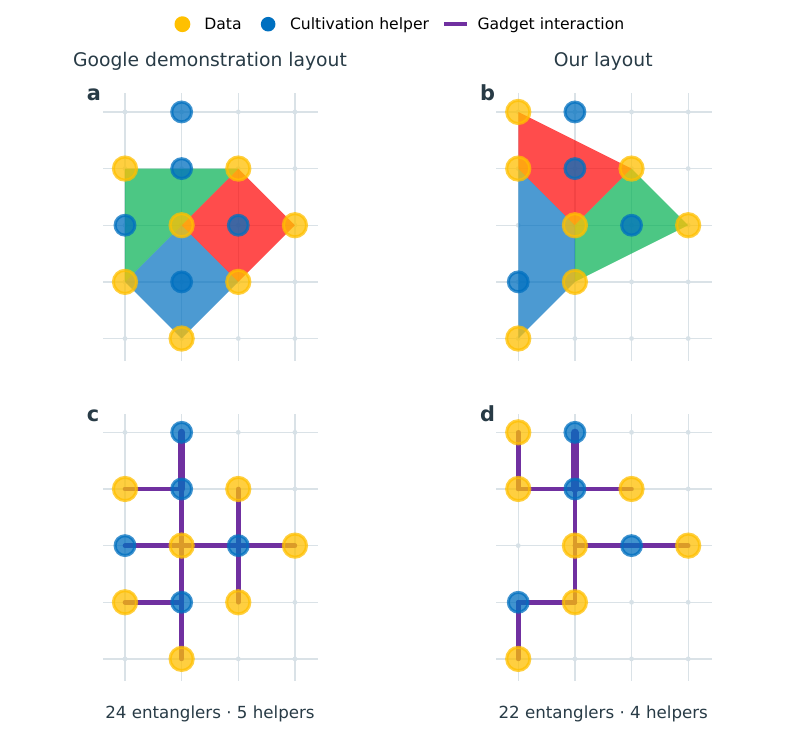}
\caption{Cultivation layouts on the QPU and their interaction costs. (a) Layout used in the Google demonstration. (b) Our layout with fixed data sites shared by injection and lattice surgery. (c) The published gadget uses 24 CNOTs and five cultivation ancilla. (d) Our gadget uses 22 CNOTs and four ancilla. Gold marks data qubits, blue marks cultivation ancilla, and colored faces show stabilizer supports.}
\label{fig:layouts}
\end{figure}

We choose a stationary layout compatible with native $R_{ZZ}$ injection and the color code surgery demonstration of Lacroix et al.~\cite{lacroix2025scaling}. It differs from the cultivation arrangement of Rosenfeld et al.~\cite{rosenfeld2026cultivation}. Both support superdense syndrome extraction (SE). Figure~\ref{fig:layouts} compares their data and cultivation ancilla sites. The reference cultivation interaction graph has no local embedding on our chosen fixed data sites. We retain its fold and unfold structure and adapt the interaction path and ancilla arrangement to the shared layout. This gives 22 CNOTs and four cultivation ancillas, compared with 24 CNOTs and five ancillas in the reference construction, while keeping the data sites fixed through preparation and resource consumption.

Figure~\ref{fig:implemented} shows the implemented gate sequence. The initial rotations select the logical measurement axis, and the fold network collects its parity onto the root ancilla. The first readout records this parity as $m$, together with flag $f_0$. The root and flag ancillas are reset before the reverse network unfolds the state. The remaining ancilla measurements record $f_1$, $f_2$, $m'$ and $f_3$. Reversing the network restores the encoded state for subsequent checks or lattice surgery.

A passing gadget requires $f_0=f_1=f_2=f_3=0$ and agreement $m=m'$. Cultivation additionally selects $m=0$, the positive logical sign. For kickback readout, both signs are retained. Each of the two analyzers must pass its internal checks, and their reported signs must agree. Accepted negative signs are counted as errors. The same fold and unfold gadget therefore serves state preparation and measurement, with the recorded sign treated according to its role. Ideal simulations verify the accepted measurement branches for both signs.

\begingroup
\providecommand{\ket}[1]{\lvert #1\rangle}
\tikzset{wire/.style={draw=black,line width=.45pt},gate/.style={rectangle,draw=black,fill=white,text=black,font=\scriptsize,minimum width=.46cm,minimum height=.38cm,inner sep=1pt}}

\begin{figure}[!tbp]
\centering
\resizebox{\linewidth}{!}{%
\begin{tikzpicture}[x=0.48cm,y=0.52cm,>=latex]
\node[anchor=west,font=\bfseries\small] at (-1.7,1) {a\quad Fold and first readout};
\draw[wire] (0,0) -- (19.8,0);
\node[anchor=east,font=\scriptsize] at (-0.12,0) {$d_{0}$};
\draw[wire] (0,-1) -- (19.8,-1);
\node[anchor=east,font=\scriptsize] at (-0.12,-1) {$d_{1}$};
\draw[wire] (0,-2) -- (19.8,-2);
\node[anchor=east,font=\scriptsize] at (-0.12,-2) {$d_{2}$};
\draw[wire] (0,-3) -- (19.8,-3);
\node[anchor=east,font=\scriptsize] at (-0.12,-3) {$d_{3}$};
\draw[wire] (0,-4) -- (19.8,-4);
\node[anchor=east,font=\scriptsize] at (-0.12,-4) {$d_{4}$};
\draw[wire] (0,-5) -- (19.8,-5);
\node[anchor=east,font=\scriptsize] at (-0.12,-5) {$d_{5}$};
\draw[wire] (0,-6) -- (19.8,-6);
\node[anchor=east,font=\scriptsize] at (-0.12,-6) {$d_{6}$};
\draw[wire] (0,-7) -- (19.8,-7);
\node[anchor=east,font=\scriptsize] at (-0.12,-7) {$h_{0}$};
\draw[wire] (0,-8) -- (19.8,-8);
\node[anchor=east,font=\scriptsize] at (-0.12,-8) {$h_{1}$};
\draw[wire] (0,-9) -- (19.8,-9);
\node[anchor=east,font=\scriptsize] at (-0.12,-9) {$h_{2}$};
\draw[wire] (0,-10) -- (19.8,-10);
\node[anchor=east,font=\scriptsize] at (-0.12,-10) {$h_{3}$};
\node[gate] at (1,-7) {$R$};
\node[gate] at (1,-8) {$R$};
\node[gate] at (1,-9) {$R$};
\node[gate] at (1,-10) {$R$};
\node[gate] at (2,0) {$R_-$};
\node[gate] at (2,-1) {$R_-$};
\node[gate] at (2,-2) {$R_-$};
\node[gate] at (2,-3) {$R_-$};
\node[gate] at (2,-4) {$R_-$};
\node[gate] at (2,-5) {$R_-$};
\node[gate] at (2,-6) {$R_-$};
\node[gate] at (3,0) {$S^\dagger$};
\node[gate] at (3,-1) {$S^\dagger$};
\node[gate] at (3,-2) {$S^\dagger$};
\node[gate] at (3,-3) {$S^\dagger$};
\node[gate] at (3,-4) {$S^\dagger$};
\node[gate] at (3,-5) {$S^\dagger$};
\node[gate] at (3,-6) {$S^\dagger$};
\node[gate] at (4,0) {$H$};
\node[gate] at (4,-1) {$H$};
\node[gate] at (4,-2) {$H$};
\node[gate] at (4,-3) {$H$};
\node[gate] at (4,-4) {$H$};
\node[gate] at (4,-5) {$H$};
\node[gate] at (4,-6) {$H$};
\node[gate] at (4,-10) {$H$};
\draw[draw=black,line width=0.8pt] (5,-10) -- (5,-9);
\fill[black] (5,-10) circle (2pt);
\draw[draw=black,fill=white,line width=0.8pt] (5,-9) circle (3.4pt);
\draw[black,line width=0.7pt] (4.88,-9) -- (5.12,-9) (5,-9.2) -- (5,-8.8);
\draw[draw=black,line width=0.8pt] (6,-5) -- (6,-1);
\fill[black] (6,-5) circle (2pt);
\draw[draw=black,fill=white,line width=0.8pt] (6,-1) circle (3.4pt);
\draw[black,line width=0.7pt] (5.88,-1) -- (6.12,-1) (6,-1.2) -- (6,-0.8);
\draw[draw=black,line width=0.8pt] (7,-1) -- (7,-9);
\fill[black] (7,-1) circle (2pt);
\draw[draw=black,fill=white,line width=0.8pt] (7,-9) circle (3.4pt);
\draw[black,line width=0.7pt] (6.88,-9) -- (7.12,-9) (7,-9.2) -- (7,-8.8);
\draw[draw=black,line width=0.8pt] (8,-4) -- (8,-9);
\fill[black] (8,-4) circle (2pt);
\draw[draw=black,fill=white,line width=0.8pt] (8,-9) circle (3.4pt);
\draw[black,line width=0.7pt] (7.88,-9) -- (8.12,-9) (8,-9.2) -- (8,-8.8);
\draw[draw=black,line width=0.8pt] (9,-3) -- (9,-8);
\fill[black] (9,-3) circle (2pt);
\draw[draw=black,fill=white,line width=0.8pt] (9,-8) circle (3.4pt);
\draw[black,line width=0.7pt] (8.88,-8) -- (9.12,-8) (9,-8.2) -- (9,-7.8);
\draw[draw=black,line width=0.8pt] (10,-8) -- (10,-2);
\fill[black] (10,-8) circle (2pt);
\draw[draw=black,fill=white,line width=0.8pt] (10,-2) circle (3.4pt);
\draw[black,line width=0.7pt] (9.88,-2) -- (10.12,-2) (10,-2.2) -- (10,-1.8);
\draw[draw=black,line width=0.8pt] (11,-6) -- (11,-7);
\fill[black] (11,-6) circle (2pt);
\draw[draw=black,fill=white,line width=0.8pt] (11,-7) circle (3.4pt);
\draw[black,line width=0.7pt] (10.88,-7) -- (11.12,-7) (11,-7.2) -- (11,-6.8);
\draw[draw=black,line width=0.8pt] (12,-7) -- (12,0);
\fill[black] (12,-7) circle (2pt);
\draw[draw=black,fill=white,line width=0.8pt] (12,0) circle (3.4pt);
\draw[black,line width=0.7pt] (11.88,0) -- (12.12,0) (12,-0.2) -- (12,0.2);
\draw[draw=black,line width=0.8pt] (13,-2) -- (13,0);
\fill[black] (13,-2) circle (2pt);
\draw[draw=black,fill=white,line width=0.8pt] (13,0) circle (3.4pt);
\draw[black,line width=0.7pt] (12.88,0) -- (13.12,0) (13,-0.2) -- (13,0.2);
\draw[draw=black,line width=0.8pt] (14,0) -- (14,-9);
\fill[black] (14,0) circle (2pt);
\draw[draw=black,fill=white,line width=0.8pt] (14,-9) circle (3.4pt);
\draw[black,line width=0.7pt] (13.88,-9) -- (14.12,-9) (14,-9.2) -- (14,-8.8);
\draw[draw=black,line width=0.8pt] (15,-10) -- (15,-9);
\fill[black] (15,-10) circle (2pt);
\draw[draw=black,fill=white,line width=0.8pt] (15,-9) circle (3.4pt);
\draw[black,line width=0.7pt] (14.88,-9) -- (15.12,-9) (15,-9.2) -- (15,-8.8);
\node[gate] at (16,-10) {$H$};
\node[gate] at (17,-9) {$X$};
\node[gate] at (18,-9) {$M_{m}$};
\node[gate] at (19,-10) {$M_{f_0}$};
\end{tikzpicture}}
\par\vspace{3mm}
\resizebox{\linewidth}{!}{%
\begin{tikzpicture}[x=0.48cm,y=0.52cm,>=latex]
\node[anchor=west,font=\bfseries\small] at (-1.7,1) {b\quad Unfold and ancilla readout};
\draw[wire] (0,0) -- (21.8,0);
\node[anchor=east,font=\scriptsize] at (-0.12,0) {$d_{0}$};
\draw[wire] (0,-1) -- (21.8,-1);
\node[anchor=east,font=\scriptsize] at (-0.12,-1) {$d_{1}$};
\draw[wire] (0,-2) -- (21.8,-2);
\node[anchor=east,font=\scriptsize] at (-0.12,-2) {$d_{2}$};
\draw[wire] (0,-3) -- (21.8,-3);
\node[anchor=east,font=\scriptsize] at (-0.12,-3) {$d_{3}$};
\draw[wire] (0,-4) -- (21.8,-4);
\node[anchor=east,font=\scriptsize] at (-0.12,-4) {$d_{4}$};
\draw[wire] (0,-5) -- (21.8,-5);
\node[anchor=east,font=\scriptsize] at (-0.12,-5) {$d_{5}$};
\draw[wire] (0,-6) -- (21.8,-6);
\node[anchor=east,font=\scriptsize] at (-0.12,-6) {$d_{6}$};
\draw[wire] (0,-7) -- (21.8,-7);
\node[anchor=east,font=\scriptsize] at (-0.12,-7) {$h_{0}$};
\draw[wire] (0,-8) -- (21.8,-8);
\node[anchor=east,font=\scriptsize] at (-0.12,-8) {$h_{1}$};
\draw[wire] (0,-9) -- (21.8,-9);
\node[anchor=east,font=\scriptsize] at (-0.12,-9) {$h_{2}$};
\draw[wire] (0,-10) -- (21.8,-10);
\node[anchor=east,font=\scriptsize] at (-0.12,-10) {$h_{3}$};
\node[gate] at (1,-9) {$R$};
\node[gate] at (1,-10) {$R$};
\node[gate] at (2,-9) {$X$};
\node[gate] at (3,-10) {$H$};
\draw[draw=black,line width=0.8pt] (4,-10) -- (4,-9);
\fill[black] (4,-10) circle (2pt);
\draw[draw=black,fill=white,line width=0.8pt] (4,-9) circle (3.4pt);
\draw[black,line width=0.7pt] (3.88,-9) -- (4.12,-9) (4,-9.2) -- (4,-8.8);
\draw[draw=black,line width=0.8pt] (5,0) -- (5,-9);
\fill[black] (5,0) circle (2pt);
\draw[draw=black,fill=white,line width=0.8pt] (5,-9) circle (3.4pt);
\draw[black,line width=0.7pt] (4.88,-9) -- (5.12,-9) (5,-9.2) -- (5,-8.8);
\draw[draw=black,line width=0.8pt] (6,-2) -- (6,0);
\fill[black] (6,-2) circle (2pt);
\draw[draw=black,fill=white,line width=0.8pt] (6,0) circle (3.4pt);
\draw[black,line width=0.7pt] (5.88,0) -- (6.12,0) (6,-0.2) -- (6,0.2);
\draw[draw=black,line width=0.8pt] (7,-7) -- (7,0);
\fill[black] (7,-7) circle (2pt);
\draw[draw=black,fill=white,line width=0.8pt] (7,0) circle (3.4pt);
\draw[black,line width=0.7pt] (6.88,0) -- (7.12,0) (7,-0.2) -- (7,0.2);
\draw[draw=black,line width=0.8pt] (8,-6) -- (8,-7);
\fill[black] (8,-6) circle (2pt);
\draw[draw=black,fill=white,line width=0.8pt] (8,-7) circle (3.4pt);
\draw[black,line width=0.7pt] (7.88,-7) -- (8.12,-7) (8,-7.2) -- (8,-6.8);
\draw[draw=black,line width=0.8pt] (9,-8) -- (9,-2);
\fill[black] (9,-8) circle (2pt);
\draw[draw=black,fill=white,line width=0.8pt] (9,-2) circle (3.4pt);
\draw[black,line width=0.7pt] (8.88,-2) -- (9.12,-2) (9,-2.2) -- (9,-1.8);
\draw[draw=black,line width=0.8pt] (10,-3) -- (10,-8);
\fill[black] (10,-3) circle (2pt);
\draw[draw=black,fill=white,line width=0.8pt] (10,-8) circle (3.4pt);
\draw[black,line width=0.7pt] (9.88,-8) -- (10.12,-8) (10,-8.2) -- (10,-7.8);
\draw[draw=black,line width=0.8pt] (11,-4) -- (11,-9);
\fill[black] (11,-4) circle (2pt);
\draw[draw=black,fill=white,line width=0.8pt] (11,-9) circle (3.4pt);
\draw[black,line width=0.7pt] (10.88,-9) -- (11.12,-9) (11,-9.2) -- (11,-8.8);
\draw[draw=black,line width=0.8pt] (12,-1) -- (12,-9);
\fill[black] (12,-1) circle (2pt);
\draw[draw=black,fill=white,line width=0.8pt] (12,-9) circle (3.4pt);
\draw[black,line width=0.7pt] (11.88,-9) -- (12.12,-9) (12,-9.2) -- (12,-8.8);
\draw[draw=black,line width=0.8pt] (13,-5) -- (13,-1);
\fill[black] (13,-5) circle (2pt);
\draw[draw=black,fill=white,line width=0.8pt] (13,-1) circle (3.4pt);
\draw[black,line width=0.7pt] (12.88,-1) -- (13.12,-1) (13,-1.2) -- (13,-0.8);
\draw[draw=black,line width=0.8pt] (14,-10) -- (14,-9);
\fill[black] (14,-10) circle (2pt);
\draw[draw=black,fill=white,line width=0.8pt] (14,-9) circle (3.4pt);
\draw[black,line width=0.7pt] (13.88,-9) -- (14.12,-9) (14,-9.2) -- (14,-8.8);
\node[gate] at (15,-10) {$H$};
\node[gate] at (15,0) {$H$};
\node[gate] at (15,-1) {$H$};
\node[gate] at (15,-2) {$H$};
\node[gate] at (15,-3) {$H$};
\node[gate] at (15,-4) {$H$};
\node[gate] at (15,-5) {$H$};
\node[gate] at (15,-6) {$H$};
\node[gate] at (16,0) {$S$};
\node[gate] at (16,-1) {$S$};
\node[gate] at (16,-2) {$S$};
\node[gate] at (16,-3) {$S$};
\node[gate] at (16,-4) {$S$};
\node[gate] at (16,-5) {$S$};
\node[gate] at (16,-6) {$S$};
\node[gate] at (17,0) {$R_+$};
\node[gate] at (17,-1) {$R_+$};
\node[gate] at (17,-2) {$R_+$};
\node[gate] at (17,-3) {$R_+$};
\node[gate] at (17,-4) {$R_+$};
\node[gate] at (17,-5) {$R_+$};
\node[gate] at (17,-6) {$R_+$};
\node[gate] at (18,-7) {$M_{f_1}$};
\node[gate] at (19,-8) {$M_{f_2}$};
\node[gate] at (20,-9) {$M_{m'}$};
\node[gate] at (21,-10) {$M_{f_3}$};
\end{tikzpicture}}
\caption{Implemented cultivation gadget and kickback protocol. (a) Fold circuit and first readout. (b) Unfold circuit and ancilla readout. Both panels show the source gate sequence on seven data qubits and four cultivation ancillas. $R$ prepares $\ket{0}$, $R_\pm=R_Z(\pm\pi/4)$ and $M$ is a $Z$ measurement. Readout labels define the recorded bits. Each gadget accepts $f_0=f_1=f_2=f_3=0$ and $m=m'$. Cultivation additionally selects $m=0$. The fold/unfold pair contains 22 CNOTs; native compilation expresses these through CZ and single qubit gates. Kickback applies two copies of this gadget after cultivation, retaining both signs and requiring their agreement. One SE round before, between, and after the analyzers supplies additional checks.}
\label{fig:implemented}
\end{figure}
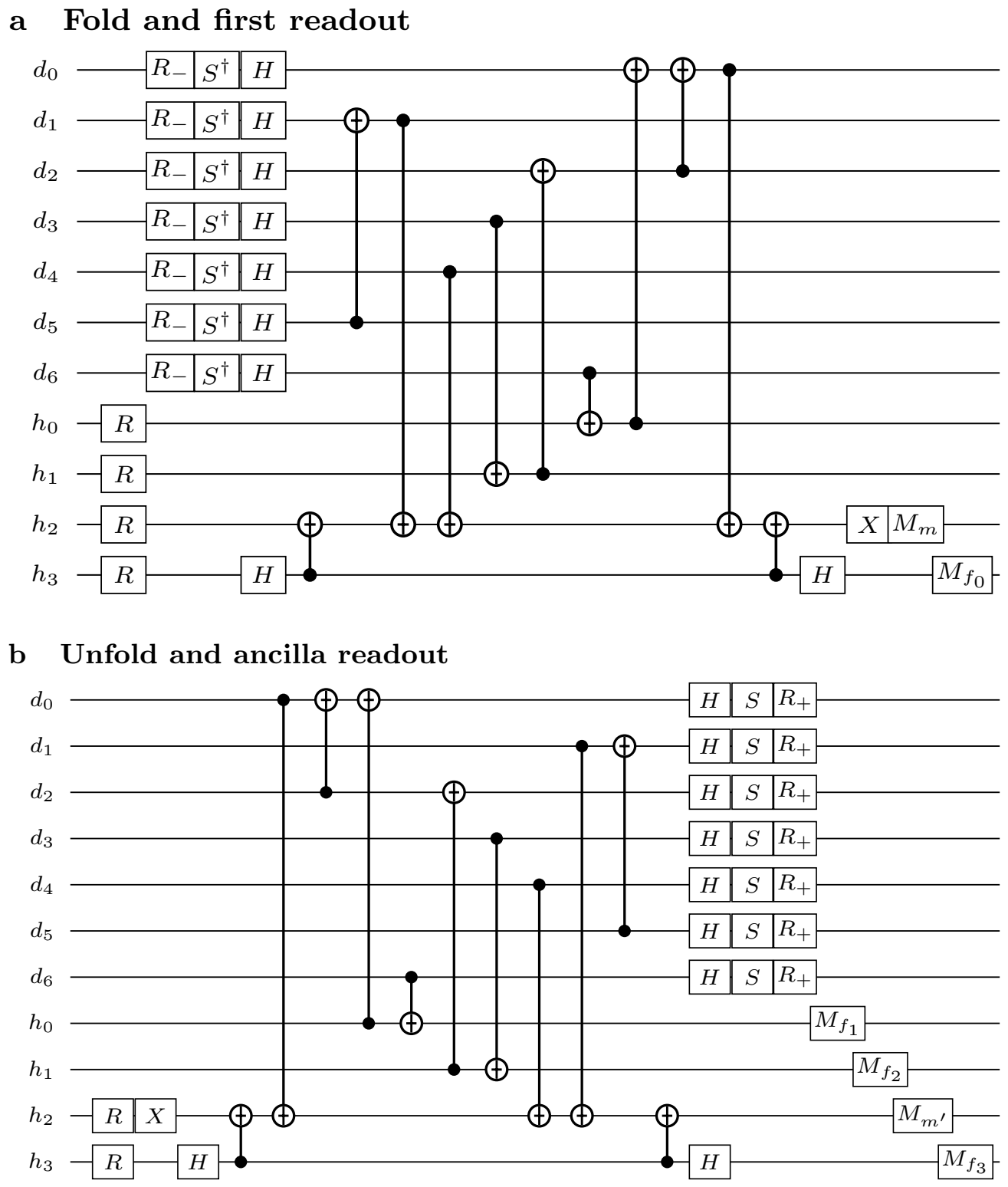
\endgroup

\section{Kickback Validation and Analyzer Bias}
\label{sec:kickback_validation}

The full kickback experiment applies two analyzers after cultivation. One SE round occurs before the first analyzer, one between the analyzers and one after the second analyzer. The terminal code checks also enter selection. These checkpoints test consistency as the state passes through the measurement sequence. We retain this complete sequence when evaluating faults and simulating the measured error probability.

We audit circuit-level Pauli faults after native gates, resets, and delays, together with classical readout flips. A single qubit location has $X$, $Y$ and $Z$ alternatives, and a two-qubit gate has all 15 non-identity Pauli alternatives. All prescribed flags, SE checks, internal analyzer agreements and agreement between analyzers remain in selection. Exact accepted wrong sign probabilities vanish for every enumerated single fault. An accepted two-fault mechanism is found both for the complete experiment and for the analyzer acting on an ideal resource.

For a representative analyzer mechanism, either fault alone fires the same SE check and makes the analyzer signs disagree. With both faults present, these detector changes cancel and both signs become negative. The internal repeated readouts of each gadget also agree, so the pair passes every check. Direct insertion into the native circuit confirms this behavior. The minimum accepted wrong sign mechanism therefore has weight two in the tested circuit-level Pauli model. An additional SE checkpoint retains two fault mechanisms.

We quantify analyzer bias by comparing noisy kickback with an ideal measurement of the same noisy prepared state. At the boundary after noisy preparation and before the first analyzer SE round, the diagnostic replaces the remaining measurement sequence by ideal operations. It resets the ancillas, projects the six code stabilizers into their positive sector and measures the logical magic state axis. This gives an input error $\epsilon_{\mathrm{in}}$ conditioned on that code projection. The full noisy analyzer gives an accepted negative sign probability $p_{\mathrm{kb}}$. Their difference, $\Delta=p_{\mathrm{kb}}-\epsilon_{\mathrm{in}}$, measures net analyzer bias for the simulated preparation and selection. Positive $\Delta$ means that kickback reports a larger error than the ideal diagnostic.

\begin{figure*}[!tbp]
\centering
\includegraphics[width=0.8 \linewidth]{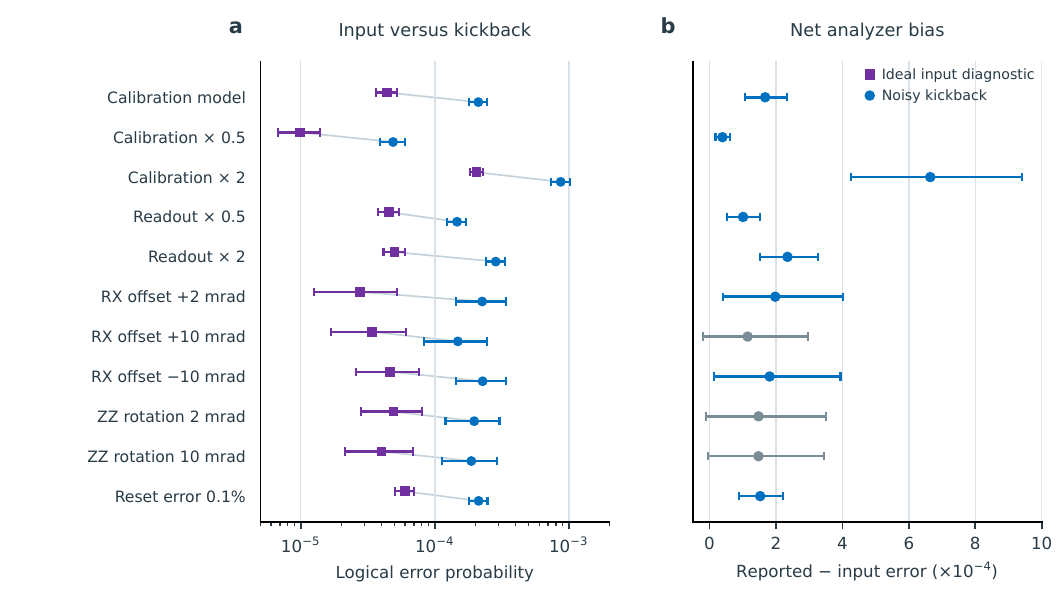}
\caption{Robustness of kickback tomography. (a) Ideal diagnostic of the prepared state and noisy kickback error, with individual 95\% binomial intervals. (b) Net reported error with at least 95\% simultaneous shot coverage across all scenarios. Blue intervals are positive; gray intervals include zero.}
\label{fig:robustness}
\end{figure*}

Fig.~\ref{fig:robustness} compares these quantities across eleven noise scenarios and 53 million simulated shots. The stochastic scenarios scale all calibrated channels, or readout alone, by $0.5$ and $2$. Coherent scenarios add fixed $R_X$ offsets of $+2$, $+10$ or $-10$ milliradians per simulated $R_X$ rotation, or $ZZ$ rotations of $2$ or $10$ milliradians after each CZ. A reset scenario adds an $X$ error with probability $0.001$ after reset. These specified perturbations test the analyzer response around the nominal calibration model. At nominal calibration, $\epsilon_{\mathrm{in}}=4.38\times10^{-5}$ and $p_{\mathrm{kb}}=2.11\times10^{-4}$. All eleven point estimates have positive net bias. Eight simultaneous confidence intervals resolve positive bias, while three include zero. We construct the simultaneous intervals with a Bonferroni allocation over both diagnostic arms and all eleven scenarios. For each scenario, subtracting the upper input error limit from the lower kickback error limit gives the lower bias limit; the opposite endpoints give the upper limit. This provides at least $95\%$ simultaneous shot coverage across the tested scenarios.

The QPU measurement gives correct sign probability $0.999859$, with $95\%$ binomial interval $[0.999741,0.999933]$, at $14.212\%$ acceptance. Across the tested noise scenarios, the simulations support interpreting this measurement as a conservative estimate of post-selected logical state fidelity. The comparison connects the observed analyzer outcome to the quality of the prepared resource through the explicitly simulated measurement sequence.

\section{Calibration Model and Offline Decoding}
\label{sec:supplement_decoding}

The calibration replay uses each circuit's saved native gates, calibration snapshot, and scheduled idle intervals. For a $d$-dimensional gate with average infidelity $r$, we use a stochastic Pauli event probability $p=(d+1)r/d$, distributed equally over the $d^2-1$ non-identity Pauli operators. This gives $p=3r/2$ for single-qubit gates and $p=5r/4$ for two-qubit gates. Virtual $R_Z$ operations are ideal. Dynamical decoupling pulses carry their calibrated gate errors, and reset is ideal in the baseline model.

For an idle interval time $t$, the Pauli twirl of relaxation and dephasing gives
\begin{equation}
p_X=p_Y=\frac{1-e^{-t/T_1}}{4},\quad
p_Z=\frac{1-2e^{-t/T_2}+e^{-t/T_1}}{4}.
\end{equation}
Here $T_1$ and $T_2$ are the saved relaxation and coherence times for the qubit. Measurement outcomes are flipped with the calibrated assignment error for the corresponding readout instruction, including fast readout. The four source groups are enabled in all 16 combinations for the Shapley budgets described in the main text. For simulated real-time corrections, the recorded noisy measurement bits determine the conditional branch, so assignment errors also affect which correction is executed.

The detector model records the syndrome changes and flips of three logical outcomes: the joint parity $p$, the resource readout $q$ and the final data parity. Fault signatures are verified by Stim propagation through the native Clifford gates. The detector relations use code constraints shared by all six inputs. Initially random measurements enter consistency relations or the tracked frame rather than being selected to a fixed sign.

Concatenated MWPM uses color-specific matching stages with CSS marginalization and graph approximations. Tesseract searches the joint detector error model, retaining correlated mechanisms and hyperedges. Its configuration uses detector beam 10, six detector orders with seed zero, beam climbing and a queue limit of $10{,}000$. Returned recoveries are checked against the measured syndrome. Both decoders infer the logical flips used to interpret $p$, $q$ and the output parity. The displayed comparison uses calibration weights for both methods, fixed independently of the test outcomes.

All four analysis arms begin with the prescribed preparation selection. Uncorrected reconstruction applies the nominal virtual teleportation corrections. The two decoder arms additionally infer a recovery while retaining surgery syndromes and flags of either value. Strict post-selection instead requires the prescribed surgery checks to pass. All arms retain both teleportation branches. Thus, the decoder and uncorrected arms use the same shots, while strict post-selection selects a smaller ensemble. Decoding is performed offline using the recorded measurements.

Each one-round tomography setting was acquired in two jobs. Earlier jobs supply the development data; the second job for every input and measurement basis supplies the held-out test data. The decoder comparison and process matrices use this common test population, including all six logical inputs. The input state panels in the main text use the full acquisition. This preserves a common comparison population for the decoder arms while using all available shots for the separate input state estimates.

\begin{figure*}[tbp]
\centering
\includegraphics[width=0.9 \linewidth]{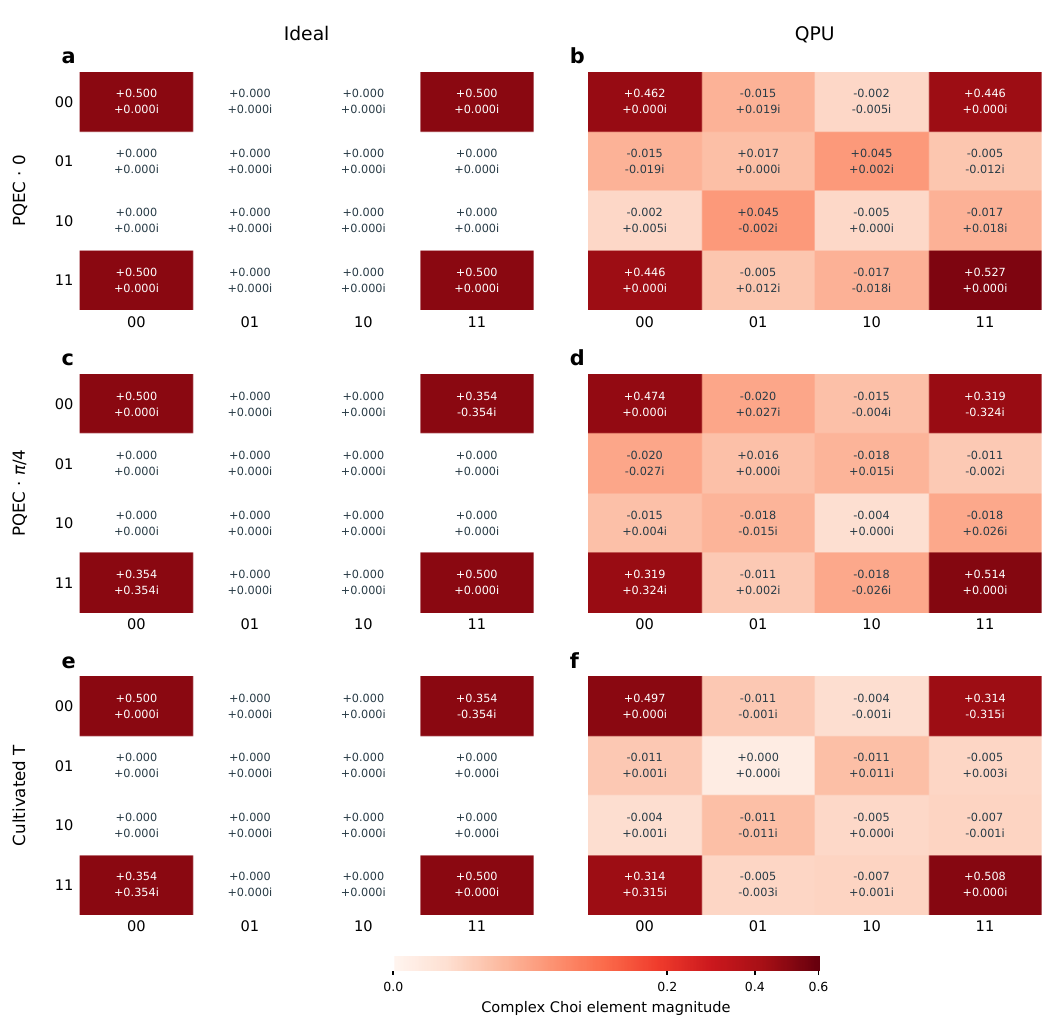}
\caption{Logical process matrices after one surgery round with full prescribed post-selection. (a) Ideal PQEC operation at $\theta=0$. (b) Corresponding QPU reconstruction. (c) Ideal PQEC operation at $\theta=\pi/4$. (d) Corresponding QPU reconstruction. (e) Ideal $T$ teleportation. (f) Corresponding QPU reconstruction with a cultivated resource. Each cell gives the real and imaginary parts on separate lines. Red intensity denotes complex magnitude on a shared power scale that emphasizes small entries; zero is white. Matrix indices are $|00\rangle$, $|01\rangle$, $|10\rangle$, and $|11\rangle$ in output and input order. Matrices are normalized to unit trace.}
\label{fig:postselected_process_matrices}
\end{figure*}

\section{Post-selected Process Matrices}
\label{sec:postselected_process_matrices}

The process matrices resolve the population and coherence underlying the reported process overlap. Figure~\ref{fig:postselected_process_matrices} compares the ideal operation with the QPU reconstruction after one surgery round, selecting all prescribed deterministic preparation and surgery checks. Both teleportation branches are retained with virtual corrections. These are the held-out post-selected results from the decoder comparison, with process fidelities of $94.08\%$ for PQEC at $\theta=0$, $94.84\%$ for partial-FT at $\theta=\pi/4$, and $94.75\%$ for cultivated $T$ teleportation.

The reconstruction uses the six cultivated Pauli eigenstates as inputs and logical $X_L$, $Y_L$, and $Z_L$ readout. For each input and output basis, accepted outcome sums are divided by the total number of raw attempts for that setting. We apply the virtual frame interpretation to these signed sums and retain the corresponding acceptance fractions. Linear inversion then reconstructs the unnormalized process from all six input Bloch vectors. We normalize its Choi matrix to unit trace only after this reconstruction. This retains input and basis dependent acceptance in the fitted process; its Choi trace gives the mean acceptance. Its overlap with the ideal normalized Choi matrix gives the process fidelity reported in the main text. The displayed matrices use strict postselection on the common test population described above. Process error bars propagate raw shot acceptance and outcome covariance through the reconstruction and trace normalization.

The basis is ordered as output and input, $|00\rangle$, $|01\rangle$, $|10\rangle$, and $|11\rangle$. Diagonal entries show populations of the Choi state, while off-diagonal entries resolve coherence and phase. At zero angle, the ideal operation is the identity. The $\pi/4$ Partial-FT resource and cultivated $T$ resource implement the same target rotation, allowing their measured matrices to be compared against the same ideal process. The matrices show how the population and complex coherence combine into the reported process overlap.

\end{document}